\documentclass[11pt,a4paper]{article}
\usepackage{jheppub}

\usepackage{epsfig,latexsym,amsfonts,amsmath,amsthm,amssymb,amsbsy,multirow,slashed,color,mathrsfs,wasysym,textcomp,subfigure,wrapfig,comment}
\usepackage{bm}

\usepackage{amsmath}
\usepackage{amssymb,amsfonts}
\usepackage{mdframed}
\usepackage{lipsum}
\usepackage{stmaryrd}

\newcommand{\nn}{\nonumber}

\newcommand{\beq}{\begin{equation}}
\newcommand{\eeq}{\end{equation}}

\newcommand{\bea}{\begin{eqnarray}}
\newcommand{\eea}{\end{eqnarray}}

\newcommand{\bal}{\begin{align}}
\newcommand{\eal}{\end{align}}

\newcommand{\ball}{\begin{align*}}
\newcommand{\eall}{\end{align*}}

\newcommand{\bes}{\begin{subequations}}
\newcommand{\ees}{\end{subequations}}

\newcommand{\uab}[1]{#1^{\alpha \beta}}

\newcommand{\dab}[1]{#1_{\alpha \beta}}
\newcommand{\dba}[1]{#1_{\beta\alpha}}
\newcommand{\brab}[1]{#1^{[\alpha \beta}}
\newcommand{\cdbr}[1]{#1^{\gamma \delta]}}
\newcommand{\asym}[2]{#1_{[\alpha \beta} #2_{\gamma \delta]}}
\newcommand{\alg}[2]{\llbracket #1 , #2 \rrbracket}

\newcommand{\inv}[1]{#1^{-1}}

\newcommand{\dg}[2]{\begin{pmatrix}
#1 & 0 \\
0 & #2 
\end{pmatrix}}

\newcommand{\sx}{$E_{6(6)}\,$}
\newcommand{\ed}{$E_{d(d)}\,$}
\newcommand{\usp}{$USp(8)\,$}
\newcommand{\rt}{\sqrt{2}}
\newcommand{\sls}{$SL(6) \times SL(2) \, $}
\newcommand{\ef}{e^{\phi}}

\newcommand{\coef}{\Big(-\dfrac{ \rho_0 e^{2A}}{4m}\Big)}

\newcommand{\of}{\omega}

\newcommand{\nc}{\mathcal{N}}

\newcommand{\kc}{\mathcal{K}}
\newcommand{\kcr}{\tilde{\mathcal{K}}} 
\newcommand{\jc}{\mathcal{J}}
\newcommand{\jcr}{\tilde{\mathcal{J}}} 

\newcommand{\D}{v}

\newcommand{\ac}{\mathcal{A}}

\newcommand{\mc}{\mathcal{M}}

\newcommand{\ML}{$\mathcal{M} \,$}

\newcommand{\rep}[1]{$\mathbf{#1}$}
\newcommand{\Rep}[1]{\mathbf{#1}}

\newcommand{\CHS}{\Gamma_{7}}

\newcommand{\tr}{\text{Tr}}
\newcommand{\cl}[1]{\dfrac{1}{#1}}
\newcommand{\ad}{$AdS_5$}
\newcommand{\nm}{\nabla_m}

\newcommand{\mt}{C} 
\newcommand{\mtf}{G} 
\newcommand{\mm}{M} 

\title{Generalized geometric vacua with eight supercharges}

 \author{Mariana Gra\~na and}  \author{Praxitelis Ntokos}

 \affiliation{Institut de Physique Th\'eorique, CEA
 Saclay, CNRS URA 2306 \\ F-91191 Gif-sur-Yvette, France} 

\emailAdd{mariana.grana@cea.fr}\emailAdd{praxitelis.ntokos@cea.fr}

\abstract{We investigate compactifications of type II and M-theory down to $AdS_5$ with generic fluxes that preserve eight supercharges, in the framework of Exceptional Generalized Geometry. The geometric data and gauge fields on the internal manifold are encoded in a pair of generalized structures corresponding to the vector and hyper-multiplets of the reduced five-dimensional supergravity. Supersymmetry translates into integrability conditions for these structures, generalizing, in the case of type IIB, the Sasaki-Einstein conditions. We show that the ten and eleven-dimensional type IIB and M-theory Killing-spinor equations specialized to a warped $AdS_5$ background  imply the generalized integrability conditions.  }

\preprint{IPhT-t16/034}

\begin{document}
\maketitle

\setcounter{footnote}{0}
\setcounter{figure}{0}
\setcounter{equation}{0}

\newpage

\newpage

\section{Introduction}

Flux compactifications play a central role both in the construction of phenomenologically-relevant models due to their potential to stabilize moduli, as well as in gauge/gravity duality where they realize duals of less symmetric gauge theories. There has been significant progress in understanding the geometry of the internal manifolds arising in flux compactifications, using the tool of G-structures, and their extension to generalized geometry.  

For the  lower dimensional  effective theory to be supersymmetric, the existence of globally defined spinors on the internal manifold is required \cite{Gauntlett:2003cy}.
This leads to a reduction of the structure group on the $d$-dimensional tangent bundle to a subgroup $G\subset SO(d)$, or in other words to have a $G$-structure. The degrees of freedom of the internal metric are then parameterized by geometric structures which are singlets of the corresponding $G$-structure. 

In generalized geometry, the metric degrees of freedom are combined with those of the gauge fields into a generalized metric. Similarly, the lower dimensional effective theory is supersymmetric if the generalized metric is encoded in structures which are singlets of a generalized $G$-structure \cite{Hitchin:2004ut,Gualtieri:2003dx,Pacheco:2008ps,Grana:2009im,Coimbra:2014uxa}. The group $G$ in this case corresponds to the structure group of the generalized tangent bundle. The latter combines the tangent bundle of the manifold, where the vectors generating the diffeomorphism symmetry of general relativity live, with powers of the cotangent bundle, whose sections are the p-forms generating the gauge symmetry of the supergravity gauge fields.    

While supersymmetric Minkowski backgrounds in the absence of fluxes are described by integrable $G$-structures, their flux analogues are integrable generalized $G$ structures \cite{Grana:2005tf,Pacheco:2008ps,Grana:2011nb,Lust:2010by,Coimbra:2015nha,Coimbra:2014uxa,Ashmore:2015joa}. This geometric reformulation of backgrounds with fluxes gives a characterization that allows in principle to find new solutions, as well as to understand the deformations, which are the moduli of the lower dimensional theory.  In the context of gauge/gravity duality, deformations of the background correspond to deformations of the dual gauge theory.
For compactifications to $AdS$, the $G$ structures are weakly integrable, and so are the corresponding generalized structures \cite{Grana:2005sn,Coimbra:2015nha,Ashmore:2016qvs}. 

In this paper we focus on $AdS_5$ compactifications of type IIB and M-theory preserving eight supercharges. These are dual to four-dimensional ${\cal N}=1$ conformal field theories. The internal manifolds are respectively five and six-dimensional. The generalized tangent bundle combines the tangent bundle plus in the case of M-theory the bundle of two and five-forms, corresponding to the gauge symmetries of the three form field and its dual six-form field, while in type IIB two copies of the cotangent bundle and the bundle of five forms and the bundle of three-forms, corresponding respectively to the symmetries of the B-field and RR 2-form field and their dual six-forms and the RR 4-form. In both cases the generalized bundle transforms in the fundamental representation of \sx, the U-duality group that mixes these symmetries.     

Compactifications leading to backgrounds with eight supercharges in the language of (exceptional) generalized geometry are characterized \cite{Grana:2009im} by two generalized geometric structures that describe the hypermultiplet and vector multiplet structures of the lower dimensional supergravity theory. When this theory is five-dimensional, the generalized tangent bundle has reduced structure group $USp(6)\subset USp(8) \subset$ \sx \cite{Ashmore:2015joa}, where $USp(8)$, the maximal compact subgroup of \sx, is the generalized analogue of $SO(6)$, namely the structure group of the generalized tangent bundle equipped with a metric. 

The integrability conditions on these structures required by supersymmetry  were formulated in \cite{Ashmore:2016qvs}. 
The ``vector multiplet" structure is required to be generalized Killing, namely the generalized vector corresponding to this structure generates 
generalized diffeomorphisms (combinations of diffeomorphisms and gauge transformations) that leave the generalized metric invariant. The integrability condition for the hypermultiplet structure requires the moment maps for generic generalized diffeomorphisms to take a fixed value proportional to the cosmological constant of $AdS$. These conditions can be seen as a generalization of Sasaki-Einstein conditions: they imply that the generalized Ricci tensor is proportional to the generalized metric. They parallel the supersymmetry conditions obtained from five-dimensional gauged supergravity \cite{Louis:2012ux}. 

In this paper, we prove the integrability conditions for the generalized structures directly from the supersymmetry equations of type IIB and eleven dimensional supergravity. For that, the generalized structures are written in terms of $USp(8)$ bispinors. These are subject to differential and algebraic conditions coming from the supersymmetry transformation of the internal and external gravitino (plus dilatino in the case of type IIB). We show that the latter imply the integrability conditions for the generalized structure.

The paper is organized as follows. Section \ref{2} starts with a short review of generalized geometry for type IIB compactifications, focusing on the case of \sx structure group relevant to compactifications down to five dimensions. We then present the generalized structures describing backgrounds with eight supercharges, and their integrability conditions for $AdS_5$ compactifications. In section \ref{3} we show that the Killing spinor equations imply the integrability conditions. We outline the key points in the main text, while leaving the details to the Appendices.     
In section \ref{4} we show the analogous statements for M-theory. Section \ref{sec:Discussion} is a short discussion of the results.

\section{Generalizing the Geometry}
\label{2}

We begin with a brief review of generalized geometry, its description of backgrounds with eight supercharges and the supersymmetry conditions. 

The starting point of generalized geometry is the extension of the tangent bundle $T\mc$ of the internal manifold to a \emph{generalized tangent bundle} $E$ in such a way that the elements of this bundle generate all of the bosonic symmetries of the theory (diffeomorphisms and gauge transformations). The generalized tangent bundle transforms in a given representation of the corresponding duality group acting on the symmetries. Following the historical path, we start by discussing the $O(d,d)$ generalized geometry, relevant to the NS-NS sector of type II theories compactified on $d$-dimensional manifolds. We then briefly introduce \ed generalized geometry which encodes the full bosonic sector of type II theories compactified on a $(d-1)$-dimensional manifold, or M-theory on a $d$-dimensional geometry. In this paper we will concentrate on the case $d=6$, i.e. compactifications of type II (in particular type IIB) and M-theory down to five dimensions, but most of  the statements in the next section are valid for any $d$.

\subsection{Geometrizing the supergravity degrees of freedom}

The NS-NS sector of type II supergravity contains the metric $g_{(mn)}$, the Kalb-Ramond field $B_{[mn]}$ and the dilaton $\phi$. The symmetries of this theory are diffeomorphisms generated by vectors $k$ and gauge transformations of the B-field which leave the $H=dB$ invariant and which are parametrized by one-forms $\of$. The combined action of these symmetries can be thought to be generated by a single object
\beq
 V=(k, \of) \ , \qquad k \in T\mc \ , \ \of \in T^*\mc
 \eeq
on the combined bundle $T\mc \oplus T^*\mc$. In fact, $V$ is well-defined only in a patch of $\mc$. If there is H-flux, in order to construct a global section of the  bundle, we need to consider 
\beq
e^B V \equiv (k, \of + \iota_k B)
\eeq
taking thus into account the non-trivial transformation of the B-field on the overlap of two patches. These \emph{generalized vectors} belong to the  \emph{generalized tangent bundle}
\beq
E \simeq T\mc \oplus T^*\mc
\eeq
where the isomorphism is provided by the $e^B$ defined above. The structure group of this bundle can be reduced from $GL(2d)$ to $O(d,d)$\footnote{The group $O(d,d)$ corresponds to the T-duality group of the massless sector of type II string theory when compactified on a $d$-dimensional manifold.} by observing that there exists an invariant metric defined by
\beq
 \eta(V, V') \equiv \cl{2}(\iota_k \of' + \iota_{k'} \of ) \ . 
\eeq

It is possible to extend many of the concepts of  ordinary differential geometry on $T\mc$ to analogues on $E$. The resulting geometry is called generalized complex geometry or $O(d,d)$-generalized geometry\footnote{For a more complete introduction to this with a focus on supergravity applications, see \cite{Coimbra:2011nw}}. 

One of the key elements in this construction is the analogue of the Lie derivative. This is the so-called \emph{Dorfman derivative} along a generalized vector $V$  on another generalized vector $V'$\footnote{By the Leibniz rule, it can be extended to arbitrary tensors constructed from $E$ and $E^*$.}. It expresses the infinitesimal action of the symmetries encoded in $V$ and is given by\footnote{Note that $V$ and $V'$ now are sections of $E$ and therefore the Dorfman derivative takes into account the non-triviality of the B-field patching.}
\beq \label{GL1}
\mathbb{L}_V V' = (\mathcal{L}_k k' , \mathcal{L}_k \of ' - \iota_{k'} d \of)
\eeq
where $\mathcal{L}$ is the ordinary Lie derivative. One can write this in a more $O(d,d)$-covariant way by embedding the ordinary derivative in a $O(d,d)$-covariant object through
\beq \label{der}
D_M = (\partial_m , 0) \in E^*
\eeq 
where $m=1,...,d$, while $M=1,...,2d$. The Dorfman or generalized Lie derivative (\ref{GL1}) takes the form
\beq \label{GL}
\mathbb{L}_V V' =( V \cdot D ) V' - (D \times  V) V'
\eeq
where $ \cdot$  and $\times$ stand respectively for  the inner product and the projection to the adjoint representation between the vector and dual vector representations\footnote{Using explicit indices, $V \cdot D=V^M D_M$, $(D \times  V)=D_M V^N |_{\rm{adjoint}}$. In the $O(d,d)$ case, the latter is  
$(D \times  V)_M{}^N=D_M  V^N- \eta^{NP} \eta_{MQ} D_P V^Q$.}.

In order to include the gauge transformations of the RR fields, or to do a generalized geometry for M-theory, one needs to extend the tangent bundle even further. Not surprisingly, the appropriate  generalized bundle should transform covariantly under the group \ed \cite{Hull:2007zu,Pacheco:2008ps}, which is the U-duality group of the massless sector of type II string theory (M-theory) when compactified on a d-1 (d) dimensional manifold. In this paper, we will deal with compactifications of type IIB and M-theory down to five dimensions, and the relevant group is therefore \sx. This extended version of generalized geometry is called  \emph{Exceptional Generalized Geometry} \cite{Coimbra:2011ky,Coimbra:2012af}. In the following sections we concentrate on the type IIB case, while in section \ref{4} we discuss the M-theory analogue.

The generalized tangent bundle for type IIB decomposes as follows
\beq\label{e6 tangent bundle}
E\simeq T\mc \oplus (T^*\mc \oplus T^*\mc ) 
\oplus \wedge^3 T^*\mc
\oplus (\wedge^5 T^*\mc \oplus \wedge^5 T^*\mc)
\eeq
where the additional components $T^*\mc \, , \,  \wedge^3 T^* \mc$ and the two copies of $\wedge^5 T^*\mc$ correspond to the gauge transformations of $C_2$, $C_4$, $C_6$ and $B_6$, the dual of $B_2$ (one can also understand this in terms of the charges of the theory, namely D1, D3, D5 and NS5 -brane charges respectively).  In the above expression, we have grouped together terms that transform as doublets under the $SL(2, \mathbb{R})$ symmetry of type IIB supergravity. 

The isomorphism implied in \eqref{e6 tangent bundle} is given by an element $e^{\mu} \in$ \sx, $\mu \in \mathfrak{e}_{6(6)}$ which can be constructed from the gauge fields of the theory in such way that the generalized vectors are well-defined in the overlap of two patches. This is in direct analogy with the $O(d,d)$ case where the only non-trivial gauge field is the B-field. The expression for $\mu$ in our case is given below in \eqref{embedgauge}.

One can also here embed the derivative in a covariant object in $E^*$, such that its non-zero components are on $T^*\mc$. The Dorfman derivative takes the same form as in the $O(d,d)$ case, namely (\ref{GL}). For its expression in terms of the $GL(5)$ decomposition of $E$ in (\ref{e6 tangent bundle}), namely the analogue of \eqref{GL1}, see \cite{Coimbra:2011ky}.  

Finally, let us mention that a complete treatment of both $O(d,d)$ and $E_{d(d)}$ generalized geometry also includes the geometrization of the so-called \emph{trombone} symmetry (see \cite{Coimbra:2011ky} for details). This is an additional $\mathbb{R^+}$ symmetry which exists in warped compactifications of M-theory and can be understood as a combination of the scaling symmetry in the eleven-dimensional theory\footnote{The M-theory action is invariant under $
g_{MN} \to e^{2\alpha} g_{MN}, \quad \mt_{3} \to e^{3 \alpha} \mt_{3}$.} (and therefore is inherited also in type II)  and constant shifts of the warp factor in the compactified theory. We incorporate the action of this symmetry by rescaling appropriately our structures (see \eqref{V and H rescaling} below) where the appearance of the dilaton in the type IIB case reflects the fact that the dilaton can be interpreted as a contribution to the warp factor in an M-theory set-up.

\subsubsection{Particular case of \sx}

Let us now specialize to the case of \sx. The generalized tangent bundle $E$ transforms in the fundamental ${\bf 27}$ representation, whose decomposition  is given in \eqref{e6 tangent bundle}. In terms of representations of $GL(5) \times SL(2)$\footnote{Here, the $SL(2)$ symmetry is the type IIB S-duality which acts linearly on the doublet of 2-form potentials and by fractional linear transformations transformations on the axio-dilaton.}, this is 
\beq \label{27GL5}
\Rep{27}=
 (\Rep{5},\Rep{1}) \oplus  (\Rep{5},\Rep{2})  \oplus (\Rep{10}, \Rep{1}) 
\oplus (\Rep{1},\Rep{2})  \ .
\eeq
It will actually turn out to be convenient to use the \sls decomposition, where the two $SL(2)$ singlets are combined into a two-vector, while the two $SL(2)$ doublets are combined into a doublet of forms. Under \sls the fundamental (anti-fundamental)  representation $V$ ($Z$) of \sx therefore decomposes as
\begin{subequations} \label{funddecomp}
\beq
\Rep{27}=(\overline{\Rep{6}}, \Rep{2}) + (\Rep{15}, \Rep{1}), \quad V=(V^{i}_{\; a}, V^{ab})
\eeq
\beq
\overline{\Rep{27}}=(\Rep{6}, \overline{\Rep{2}}) + (\overline{\Rep{15}}, \Rep{1}), \quad Z=(Z^{a}_{\; i}, Z_{ab})
\eeq
\end{subequations}
where $a,b,c,\dots$ run from 1 to 6 and $i,j,k,\dots$ from 1 to 2. 

The derivative embeds naturally in the anti-fundamental representation as\footnote{The reason for the additional factor of $e^{2\phi/3}$ is related to the rescaling of the bispinors which will be introduced later, see \eqref{V and H rescaling}.}

\beq\label{embedder}
D^i_m=D^i_6=D_{mn}=0, \qquad D_{m6}=e^{2\phi/3}\partial_m 
\eeq
where we use $m,n, \dots$ for the coordinate indices on the internal manifold.

The adjoint representation splits under \sls as
\beq \label{adjdecomp}
\Rep{78}=(\Rep{35}, \Rep{1}) + (\Rep{1}, \Rep{3}) + (\overline{\Rep{20}}, \Rep{2}), \quad \mu = (\mu^a_{\; b},\mu^i_{\; j},  \mu^i_{abc}) \ .
\eeq
In our conventions, the dilaton and gauge fields embed in this representation in the following way
\begin{subequations}\label{embedgauge}
\beq
\mu^1_{mn6}= \ef  C_{mn}
\eeq
\beq
\mu^2_{mn6}=B_{mn}
\eeq
\beq \label{embedgaugemn}
\mu^m_{\, \, \,n}= -\dfrac{\phi}{6} \delta^m_n
\eeq
\beq \label{embedgauge66}
\mu^6_{\, \, \,6}= \dfrac{5 \phi}{6}
\eeq
\beq
\mu^n_{\;6} =  -\ef (\ast C_4)^n
\eeq
\beq
 \mu^i_{\,\,j} = \begin{pmatrix}
 & - (\phi/2) & \ef C_0 \\
 & 0 & (\phi/2)
 \end{pmatrix}
\eeq
\end{subequations}
while the other components of $\mu$ vanish\footnote{These other components of $\mu$ could have non-vanishing values in a different U-duality frame.}. Note that the the gauge fields from the RR sector carry an $\ef$ factor.

\subsection{Backgrounds with eight supercharges}
\label{Geometrizing the supersymmetry conditions}

In the previous section we mentioned briefly how the supergravity degrees of freedom can be packed into generalized geometric objects which belong to representations of the corresponding duality group.
In this section, we focus on the case of backgrounds that have eight supercharges off-shell, and in the next subsection we show how the on-shell restriction (i.e., the requirement that the background preserves the eight supercharges) is written in the language of exceptional generalized geometry.

Backgrounds with off-shell supersymmetry are characterized in ordinary geometry by the existence of well-defined spinors, or in other words a reduction of the structure group of the tangent bundle from $SO(d)$ to subgroups of it singled out by the fact that they leave the well-defined spinors invariant. This means that the metric degrees of freedom can be encoded in objects that are invariant under the structure group, built out of bilinears of the spinors. For the familiar case of $SU(d/2)$ structures (like the case of Calabi-Yau), these objects are the K\"ahler 2-form $\omega$ and the holomorphic d/2-form $\Omega$, satisfying certain compatibility conditions\footnote{\label{foot:compat}These are $ \omega \wedge \Omega=0$, $\omega^{d/2}= \dfrac{(d/2)!}{2^{d/2}} (-1)^{\frac{d (d/2+1)}{4}} i^{d/2} \Omega \wedge \bar \Omega$.}.

On-shell supersymmetry imposes differential conditions on the spinors, which are translated into differential conditions on the bilinears of spinors. In the absence of fluxes, the supersymmetric solutions involve an external Minkowski space, and the differential conditions lead to integrable structures on the internal space. In the case of M-theory compactifications down to five dimensions preserving eight supercharges, the internal manifold has to be Calabi-Yau, namely the K\"ahler 2-form and the holomorphic 3-form are closed. 

Compactifications to $AdS$ require on one hand some flux to support the curvature, and on the other hand the integrability conditions are weaker (they are usually referred to as weakly integrability conditions). For full integrability all torsion classes are zero, while for weak integrability there is a torsion in a singlet representation of the structure group, proportional to the curvature of $AdS$. The simplest example of compactifications to $AdS_5$ is that of type IIB, where the curvature is fully provided by the 5-form flux, and the internal space is Sasaki-Einstein (the simplest case being $S^5$).  Sasaki-Einstein manifolds are U(1)-fibrations over a K\"ahler-Einstein base (defined by a K\"ahler 2-form $\omega_{B}$ and a holomorphic 2-form $\Omega_{B}$ satisfying the compatibility condition) and a contact structure $\sigma$, satisfying
\beq
d \sigma = 2 m \, \omega_{B}, 
\qquad d\Omega_B= 3 i m\,  \sigma \wedge \Omega_B
\ 
\eeq
where $m$ is at the same time the curvature of the internal space (more precisely, the Einstein condition is $R_{mn}= 4 m^2 g_{mn}$), that of $AdS_5$, and give also the units of five-form flux. The integrability conditions on the structures for more general solutions were obtained in \cite{Gauntlett:2005ww}. 

In M-theory there is no such a simple $AdS_5$ solution. The most well known solution is that of Maldacena and Nu\~nez \cite{Maldacena:2000mw}, corresponding to the near horizon limit of M5-branes wrapped on holomorphic cycles of a Calabi-Yau 3-fold. More general solutions are studied in \cite{Gauntlett:2004zh}, and correspond topologically to fibrations of a two-sphere over a K\"ahler-Einstein base. 

The effective five-dimensional gauged supergravity encodes the deformations of the background. When there is a G-structure, the moduli space of metric deformations is given by the deformations of the structures. Together with the moduli coming from the B-field and the RR fields, they form, in the case of ${\cal N}=2$ gauged supergravity, the hypermultiplets and vector multiplets of the effective theory. 

In the generalized geometric language, metric degrees of freedom can also be encoded in bilinears of spinors (this time transforming under the the compact subgroup of the duality group, namely \usp for the case of \sx), and furthermore these can be combined with the degrees of freedom of the gauge fields such that the corresponding objects (called generalized structures\footnote{In the case of $O(d,d)$ generalized geometry these are $Spin(d,d)$ pure spinors.}) transform in given representations of the \ed group. 
For eight supercharges in five dimensions the relevant generalized structures form a pair of objects $(K, J_a)$,  first introduced in \cite{Grana:2009im}. In the next section we are going to give their explicit form, but for the moment let us explain their geometrical meaning.

The structure $K$ transforms in the fundamental representation of $E_{6(6)}$ and it is a singlet under the $SU(2)$ R-symmetry group of the relevant effective supergravity theory. If $K$ was to be built just as a bispinor (we will call that object $\kc$, its explicit expression is given in \eqref{K}), then it would be a section of the right-hand side of \eqref{e6 tangent bundle} and it would not capture the non-trivial structure of the flux configuration on the internal manifold. Therefore, the proper generalized vector which transforms as a section of $E$ is the dressed one 
\beq \label{dressedK}
K = e^{\mu} \kc \ .
\eeq
This structure was called the V-structure (vector-multiplet structure) in \cite{Ashmore:2015joa} since it parametrizes the scalar fields of the vector multiplets in the effective theory.

The other algebraic structure, or rather an $SU(2)_R$ triplet of structures, describing the hypermultiplets (and thus called H-structure in \cite{Ashmore:2015joa}) is $J_a, \, a=1,2,3$. It transforms in the adjoint of $E_{6(6)}$. As for $K$, we need the dressed object 
\beq
J_a = e^{\mu} {\cal J}_a e^{-\mu}= e^{ \alg{\mu}{\cdot}} \jc_a
\eeq
where we are using $\alg{\cdot}{\cdot}$ to denote the $\mathfrak{e}_{6(6)}$ adjoint  action. These are normalized as\footnote{We use the notation $\text{Tr}(\cdot ,\cdot)$ to denote the Killing form for $\mathfrak{e}_{6(6)}$.}
\beq \label{norm}
 \tr(\jc_a ,\jc_b ) = 8 \rho^2 \delta_{ab}
\eeq
where $\rho$ will be related to the warp factor, and satisfy the $SU(2)$ algebra 
\beq\label{su2algebra}
\alg{\jc_a}{\jc_b} = (4 i \rho) \epsilon_{abc} \jc_c \ .
\eeq

As in Calabi-Yau compactifications where $\omega$ and $\Omega$ have to satisfy compatibility conditions to define a proper Calabi-Yau structure (see footnote \ref{foot:compat}), similar requirements apply here, and read  
\beq\label{comp}
\jc_a \, \kc =0 \ , \qquad c(\kc,\kc,\kc)= 6 \rho^3
\eeq
where in the first expression we mean the adjoint action of $\jc$ on $\kc$, and in the second one $c$  is the cubic invariant of \sx. Since the above expressions are \sx -covariant, they have exactly the same form if we replace $(\kc, \jc_a)$ with their dressed version $(K,J_a)$.

\subsection{Supersymmetry conditions}

In the previous section we have introduced the generalized structures defining the backgrounds with eight supercharges off-shell, namely those that allow to define a five-dimensional (gauged) supergravity upon compactification. Here we discuss the integrability conditions that these backgrounds need to satisfy  
in order to preserve all eight supersymmetries leading to an $AdS_5$ geometry on the external space. The supersymmetry conditions were originally introduced in \cite{Ashmore:2016qvs}, and the relevant backgrounds called ``exceptional Sasaki-Einstein" (the simplest case corresponding to Sasaki-Einstein manifolds). Here we will write the supersymmetry conditions in a slightly different way, and in the next section we will use the fact that they are independent of the (generalized) connection to choose a convenient one to verify them directly from the 10d supersymmetry conditions.   

Compactifications to warped $AdS_5$  require, both in M-theory and in type IIB 
\beq\label{Hint}
D \tilde{J}_a + \kappa \,  \epsilon_{abc}
\tr (\tilde{J}_b , D\tilde{J}_c )=
\lambda_a   c(\tilde{K},\tilde{K}, \cdot)
\eeq
\beq\label{Vint}
\mathbb{{L}}_{\tilde{K}} \tilde{K} = 0
\eeq
\beq\label{compint}
\mathbb{{L}}_{\tilde{K}}\tilde{J}_a = \dfrac{3i}{2} \epsilon_{abc} \lambda_b \tilde{J}_c
\eeq
These equations involve the rescaled bispinors, which for type IIB are  (the analogue expressions for M-theory are given in \eqref{V and H rescaling M}) 
\beq\label{V and H rescaling}
\tilde{K} = e^{ - 2 \phi / 3} K \ , \qquad
\tilde{J}_a = e^{2A - 2 \phi} J_a \ ,
\eeq
where $A$ is the warp factor and $\phi$ the dilaton. $D$ is the derivative defined in \eqref{embedder}, whose explicit index we have omitted, and corresponds to the direction missing in the cubic invariant\footnote{To write this index explicitly we substitute $D \to D_M$, $c(\tilde{K},\tilde{K}, \cdot) \to c_{MNP} \tilde K^N \tilde K^{P}$.}. 
The coefficient $\kappa$ is related to the normalization of the structures and is given by 
\beq
\dfrac{1}{\kappa}
=i \| \tilde{J}_a \| 
\equiv i   \sqrt{8\text{Tr}(\tilde{J}_a ,\tilde{J}_a )} 
 \eeq
and for type IIB is\footnote{Note that $\kappa$ accounts for both the normalization of the internal spinors (Eq. \eqref{normequality}) and the rescalings \eqref{V and H rescaling} as can be seen by writing it as $\kappa=(8i \rho e^{2A - 2 \phi})^{-1}$.}
\beq
\kappa= - \dfrac{i}{4\rt} e^{-3A + 2 \phi} .
\eeq 
Finally, $\lambda_a$ are a triplet of constants related to the $AdS_5$ cosmological constant $m$ by
\beq\label{lambda values}
\lambda_1= \lambda_2 =0 , \quad \lambda_3 = -2 i m \ .
\eeq

Let us explain very briefly the meaning of these equations. For more details, see \cite{Ashmore:2015joa,Ashmore:2016qvs}. The first equation which one can write in terms of the Dorfman derivative along a generic generalized vector,\footnote{The expression is as follows 
$$
{\kappa} \, \epsilon_{abc} \tr\alg{ \tilde{J}_b}{ \mathbb{{L}}_V \tilde{J}_c } = \lambda_a  c(\tilde{K},\tilde{K}, V) \ .
$$} implies that the moment map for the action of a generalized diffeomorphism along $V$ takes a fixed value that involves the vector multiplet structure and the $SU(2)_R$ breaking parameters $\lambda_a$ ($AdS_5$ vacua only preserve a $U(1)_R \in SU(2)_R$ \cite{Louis:2012ux,Louis:2016qca}), given by $\lambda_a J_a$. The second and third equation imply that $\tilde K$ is a generalized Killing vector  of the background. Indeed,  (\ref{Vint}) implies that it leaves $\tilde K$ invariant, while  \eqref{compint} shows that the generalized diffeomorphism along $\tilde K$ amounts to an $SU(2)_R$ rotation of the $J_a$. This rotation does not affect the generalized metric which encodes all the bosonic degrees of freedom.  Thus, the generalized vector $\tilde K$  was called  ``generalized Reeb vector" of the exceptional Sasaki-Einstein geometry. 

As shown in \cite{Ashmore:2016qvs}, these conditions imply that these backgrounds are generalized Einstein, as the generalized Ricci tensor is proportional to the generalized metric.

We can compare these to the conditions coming from the five dimensional gauged supergravity \cite{Louis:2012ux}. More specifically, \eqref{compint} corresponds to the hyperini variation, \eqref{Vint} corresponds to the gaugini, while \eqref{Hint} corresponds to a combination of the gravitini and the gaugini.

In the next section, we will give more details of the construction of H-and V structures in terms of internal spinors, and we show by an explicit calculation that $AdS_5$ compactifications preserving eight supercharges require conditions \eqref{Hint}-\eqref{compint}.

\section{From Killing spinor equations to Exceptional Sasaki Einstein conditions}
\label{3}

\subsection{IIB compactifications to $AdS_5$ with general fluxes}

In this section we show that supersymmetry requires the integrability conditions (\ref{Hint})-\eqref{compint}.

We are interested in solutions of type IIB supergravity which
\begin{itemize}
\item
respect the isometry group $SO(4,2)$ of $AdS_5$ and
\item
preserve 1/4 of the original supersymmetry, i.e. 8 supercharges.
\end{itemize}
According to the former condition, the ten-dimensional metric is written as
\beq\label{metric ansatz}
ds^2=e^{2A(y)}\tilde{g}_{\mu\nu}(x) dx^{\mu}dx^{\nu}+g_{mn}(y) dy^m dy^n
\eeq
where $\tilde{g}_{\mu\nu}(x)$ is the metric of $AdS_5$ and $g_{mn}(y)$ is the metric of the internal manifold, while the fluxes are of the form
\beq
G_{(n)}=F_{(n)} + {\rm vol}_5 \wedge \hat F_{(n-5)} 
\eeq
where $F_{(n)}$ is purely an internal piece. 

We start with the supersymmetry transformations of type IIB supergravity for the gravitino and the dilatino which read respectively (in the democratic formulation \cite{Bergshoeff:2001pv})
\beq \label{10DG}
\delta \Psi_M=
\nabla_M \epsilon - \dfrac{1}{4} \slashed{H}_M \sigma^3 \epsilon + \dfrac{e^\phi}{16}  \left[( \slashed{G}_1 +\slashed{G}_5 + \slashed{G}_9) \Gamma_M (i\sigma^2) + ( \slashed{G}_3 +\slashed{G}_7 ) \Gamma_M \sigma^1 \right] \epsilon 
\eeq
\beq \label{10DD}
\delta \lambda=
\left( \slashed{\partial}\phi - \dfrac{1}{2} \slashed{H} \sigma^3 \right)\epsilon  - \dfrac{e^\phi}{8}  \left[4 (\slashed{G}_1 - \slashed{G}_9) (i\sigma^2) +2 (\slashed{G}_3 - \slashed{G}_7) \sigma^1 \right] \epsilon 
\eeq
where 
$\slashed{G}_n=\cl{n!} G_{M_1 \dots M_n} \hat{\Gamma}^{M_1 \dots M_n }$  (we are using hats for quantities defined in ten dimensions) and $\sigma^1,\sigma^2,\sigma^3$ are the Pauli matrices acting on the doublet of type IIB spinors
\beq
\epsilon=
\begin{pmatrix}
\epsilon_1 \\
\epsilon_2
\end{pmatrix}.
\eeq

For backgrounds preserving eight supercharges, we parametrize\footnote{Our conventions for spinors and gamma matrices as well as their properties are described in appendix \ref{Spinor conventions}.} the ten-dimensional supersymmetry parameters $\epsilon_i$ as 
\beq\label{10decompospinor}
\epsilon_i=\psi \otimes \chi_i \otimes u + \psi^c \otimes \chi_i^c \otimes u, \quad i=1,2.
\eeq
Here $\psi$ stands for a complex spinor of $Spin(4,1)$ which represents the supersymmetry parameter in the corresponding five-dimensional supergravity theory, and satisfies the Killing spinor equation of \ad 
\beq
\nabla_{\mu} \psi = \dfrac{m}{2} \rho_{\mu} \psi
\eeq
where $m$ is the curvature of the $AdS$\footnote{Five-dimensional Minkowski solutions are described by taking appropriately the limit $m \to 0$.}. $(\chi_1, \chi_2)$ is a pair of (complex) sections  of the $Spin$ bundle for the internal manifold.  The two component complex object $u$ fixes appropriately the reality and chirality properties of the ten-dimensional supersymmetry parameters $\epsilon_i$ (see \eqref{u properties}).

Inserting this decomposition in \eqref{10DG} and \eqref{10DD} and requiring the variations to vanish gives rise to 3 equations corresponding to the external gravitino, internal gravitino and dilatino respectively:\footnote{Note that for the Sasaki-Einstein case we have $\chi_2 = i\chi_1$ and in the simplest example only the five-form flux is present.}

\beq\label{egspinor}
 \left[m -e^A
 (\slashed{\partial}A) \Gamma^6 \CHS + i\dfrac{e^{\phi +A}}{4} \left((\slashed{F}_1 +\slashed{F}_5) \Gamma^6 - \slashed{F}_3\right)
  \right] 
\begin{pmatrix}
\chi_1 \\ \chi_2
\end{pmatrix}=0
\eeq

\beq\label{igspinor}
\left[\nabla_m-\dfrac{1}{4}\slashed{H}_m \Gamma^6 + i \dfrac{e^{\phi}}{8}  \left(\slashed{F}_1 +\slashed{F}_5 - \slashed{F}_3 \Gamma^6   \right)\Gamma_m \Gamma_{(7)} \right] 
\begin{pmatrix}
\chi_1 \\ \chi_2
\end{pmatrix}=0
\eeq

\beq\label{dspinor}
\left[
 (\slashed{\partial} \phi) \Gamma^6 \Gamma_{(7)} + \dfrac{1}{2} \slashed{H} \CHS - \dfrac{i e^{\phi}}{2}\left(2 \slashed{F}_1 \Gamma^6 -\slashed{F}_3 \right)
  \right]
\begin{pmatrix}
\chi_1 \\ \chi_2
\end{pmatrix}=0
\eeq
where we have used the duality ${\star}_{10} G_n =(-)^{\text{Int}[n/2]} G_{10-n} $ to write the fluxes $\hat F$ in terms of purely internal components
$F$. The $\Gamma$- matrices appearing in the above equations are constructed from the ten-dimensional ones as shown in appendix \ref{Spinor conventions}.

Now, let us mention some generic properties of IIB flux compactifications down to $AdS_5$ which are implied by the supersymmetry requirements. Although these statements can be proved without any reference to generalized geometry (as in \cite{Gauntlett:2005ww}), we will postpone their proof until appendix \ref{app:xsi} to see how nicely this formalism incorporates them. Here, we just state them.

The first property has to do with the norms of the internal spinors. From \eqref{aux2}, we see that the two internal spinors have equal norms and from \eqref{rho dependence} that they scale as $e^A$:\footnote{Note that the $\rho$ defined here is the same as the one appearing in the normalization condition of $\jc$, Eq. \eqref{norm}.} 
\beq\label{normequality}
\chi_1^{\dagger} \chi_1 = \chi_2^{\dagger} \chi_2 \equiv \rho = \dfrac{e^A}{\rt}
\eeq
Moreover, \eqref{aux3} expresses the following orthogonality property
\beq\label{orthogonality}
\chi_1^{\dagger} \chi_2 + \chi_2^{\dagger} \chi_1 = 0 
\eeq

An important consequence of the supersymmetry conditions which will be crucial for the geometrical characterization of \ML  is the existence of an isometry parametrized by a vector $\xi$ \cite{Gauntlett:2005ww}, the so-called Reeb vector\footnote{In the context of AdS/CFT, this isometry corresponds in the dual picture to the surviving R-symmetry of the $\nc=1$ gauge theory.}. The components of $\xi$ can be constructed from spinor bilinears as
\beq\label{Killingdef}
\xi^m = \cl{\rt}(\chi_1^{\dagger}\gamma^m \chi_1 + \chi_2^{\dagger}\gamma^m \chi_2)
\eeq
Actually, it turns out (see Appendix \ref{app:xsi})  that $\xi$ generates a symmetry of the full bosonic sector of the theory:
\beq\label{Killingfull}
\mathcal{L}_{\xi} \{g, A, \phi, H , F_1,  F_3 , F_5\}=0 .
\eeq
Using this, we can easily see that the Lie derivatives $\mathcal{L}_{\xi}{\chi_i}$ of the spinors satisfy the same equations \eqref{egspinor} - \eqref{dspinor} as the spinors themselves\footnote{Here, note that the existence of the isometry is crucial for the Lie derivative to commute with the covariant one.} and so they are proportional to them which means that they have definite charge. This charge is computed in appendix \ref{app:xsi}. From  \eqref{spinor charges} we have
\beq\label{spinorcharge}
\mathcal{L}_{\xi} \chi_i = \dfrac{3im}{2}  \chi_i
\eeq
These conditions are very useful in proving the integrability conditions in the next section.

\subsection{The H and V structures as bispinors}

Let us now construct the H and V structures from the internal spinors, as appropriate \sx objects. For this, it is useful to decompose the group in its maximal compact subgroup \usp.\footnote{Here, we just present some basic facts. More details are given in appendix \ref{e6reps}.}

The fundamental \rep{27} (anti-fundamental $\overline{\Rep{27}}$ ) representation is undecomposable, and corresponds to an antisymmetric $8 \times 8$ matrix $\uab{V}$ ($\uab{Z}$) which is traceless with respect to the symplectic form $C_{\alpha \beta}$ of \usp  
\beq
\Rep{27} \ ,  \quad  V=V^{\alpha\beta} \ , \quad \rm{such \ that} \ V^{\alpha \beta} C_{\alpha \beta}=0
\eeq
The adjoint \rep{78} representation corresponds to a symmetric $8 \times 8$ matrix and a fully antisymmetric rank 4 tensor
\beq\label{adj split}
\Rep{78} = \Rep{36} + \Rep{42}, \quad \mu = (\mu^{\alpha\beta}, \mu^{\alpha \beta \gamma \delta})
\eeq
The internal spinors $(\chi_1,\chi_2)$ which are sections of $Spin(5) \cong USp(4)$, are combined into the following  \usp spinors  
\beq\label{theta definition}
\theta_1=\begin{pmatrix}
\chi_1 \\ \chi_2
\end{pmatrix},
\quad
\theta_2=\begin{pmatrix}
\chi_1^c \\ \chi_2^c
\end{pmatrix}.
\eeq
In terms of the \usp spinors $\theta_i$, the normalization condition \eqref{normequality} implies
\beq\label{normusp8}
\theta^{*\alpha}_i \theta_{j,\alpha}= 2 \rho \ \delta_{ij}.
\eeq

Now, one can define the H and V structures as bispinors in a natural way. The triplet of H structures $\jc_a$ are defined as

\beq\label{defj}
(\mathcal{J}_a)_{\alpha}^{\; \beta}=(\sigma_a)^{ij} \theta_{i, \alpha} \theta_j^{\star \beta} 
\eeq
where $\sigma_a =(\sigma_1, \sigma_2, \sigma_3)$ are the Pauli matrices. Note that $\jc_a$ have components only in the \rep{36} piece of the \rep{78}.

For the V structure, we have 

\beq\label{K}
\mathcal{K}^{\alpha \beta}= \mathcal{J}_0^{\alpha \beta} - \dfrac{1}{8}C^{\alpha \beta} C_{\delta \gamma} \mathcal{J}_0^{\gamma \delta} \ , \quad {\rm with} \   \ (\mathcal{J}_0)_{\alpha}^{\; \beta}=\delta^{ij} \theta_{i, \alpha} \theta_j^{\star \beta} 
\eeq
where $C^{\alpha \beta}$ is the charge conjugation matrix, which in our conventions is the symplectic form of \usp. Note that $\kc$ is traceless by construction. From now on, we will drop the \usp indices $\alpha, \beta$ in $\kc$, $\jc$.

The $\mathfrak{su}(2)$ algebra of the structures ${\cal J}_a$, Eq. \eqref{su2algebra}, follows from the orthogonality and normalization of the spinors \eqref{normusp8}. Similarly we have 
\begin{subequations}
\beq\label{su2algebra extend}
\jc_a^2 = \jc_0^2 = 2 \rho \jc_0
\eeq
\beq\label{usp8 compatibility}
\jc_0 \jc_a = \jc_a \jc_0 =  2 \rho \jc_a
\eeq
\end{subequations}
where $\rho$ can also be related to the trace part of $\jc_0$, namely 
\beq \label{rho}
\rho= \cl{4} \tr[\jc_0] 
\eeq

The fact that $\jc_a$ and $\jc_0$ commute translates in \sx language (by using \eqref{usp8actiona}) into the compatibility condition \eqref{comp}.

In the following, it will turn out useful to have explicitly the $GL(5) \times SL(2)$ components of ${\cal K}$ and ${\cal J}_a$. For the former, using the decomposition of the $\bf{27}$ representation given in \eqref{27GL5}, we have: 
\beq
\kc= [ \xi , (\zeta , \zeta_{7}), V, (R, R_7)] \ .
\eeq
These can be organized in terms of a Clifford expansion as
\beq\label{k expansion}
\kc=\dfrac{1}{2\rt}\Big[i \xi_m \Gamma^{m67} +\zeta_m \Gamma^m +i \zeta^7_m \Gamma^{m7} 
+\dfrac{i}{2} V_{mn} \Gamma^{mn7} \Big]
\eeq
where the various components can be obtained by taking appropriate traces with $\kc$ \footnote{For example, $\xi^m=\frac{1}{2\sqrt{2}} \tr[\kc \Gamma^{m67}]$.}. In terms of bilinears involving the internal spinors $\chi_1$ and $\chi_2$ these components are 
\begin{eqnarray}\label{neutral0 bilinears}
\zeta^{m}&=&\cl{\rt}(\chi_1^{\dagger}\gamma^{m}\chi_2+\chi_2^{\dagger}\gamma^{m}\chi_1) \nn \\
\zeta_7^{m}&=&\cl{\rt}
(-\chi_1^{\dagger} \gamma^{m} \chi_1 + \chi_2^{\dagger} \gamma^{m} \chi_2) \nn \\
\xi^m&=&\cl{\rt}
(\chi_1^{\dagger} \gamma^{m} \chi_1 + \chi_2^{\dagger} \gamma^{m} \chi_2) \\
V^{mn}&=&\cl{\rt}(\chi_1^{\dagger}\gamma^{mn}\chi_2-\chi_2^{\dagger}\gamma^{mn}\chi_1) \nn \\
R&=&\cl{\rt}(\chi_1^{\dagger}\chi_1-\chi_2^{\dagger}\chi_2) \nn \\
R_7&=&\cl{\rt}(\chi_1^{\dagger}\chi_2+\chi_2^{\dagger}\chi_1) \nn 
\end{eqnarray}
Note the absence of $R$ 
and $R_7$
in the expansion \eqref{k expansion}. This is because these vanish as a consequence of the supersymmetry conditions that impose the two internal spinors to be orthogonal and have equal norm (see \eqref{normequality}, \eqref{orthogonality}). Moreover, note that the vector component $\xi$ of $\kc$ appearing in the above expression is the Reeb vector given in \eqref{Killingdef}. 

For the particular case of Sasaki-Einstein manifolds, where $\chi_2=i \chi_1$, also the one-forms  $\zeta$ and $\zeta_7$ are zero, while the two-form $V$ corresponds to $*(\sigma\wedge \omega_B)$.\footnote{The Reeb vector $\xi$ and the contact structure $\sigma$ satisfy $\iota_\xi \sigma=1$.} 
 The holomorphic 2-form of the base $\Omega_B$ is instead embedded in $J_a$, to which we now turn. 

The triplet $\jc_a$ is in the ${\bf 36}$ representation of \usp, which decomposes under $GL(5) \times SL(2)$ as
\beq\label{adj36 split}
\Rep{36}=(\Rep{5},\Rep{1})+ (\Rep{10},\Rep{1}) + (\Rep{1},\Rep{1})  +( \Rep{10}, \Rep{2}) \ .
\eeq
The Clifford expansion of ${\cal J}_a$ is \footnote{We use the notation $\jc_a^{(I)} = \tr[\jc_a \Gamma^{(I)}], \, a=1,2,3$ where $(I)$ is a collection of indices.}
\beq\label{j expansion}
\jc_a =-\dfrac{1}{8}\Big[\jc_a^{m6} \Gamma_{m6} +\cl{2}\jc_a^{mn} \Gamma_{mn} 
- \jc_a^7 \Gamma_7
+\cl{2}\jc_a^{mn6}  \Gamma_{mn6}+\cl{6}\jc_a^{mnp} \Gamma_{mnp}  \Big]
\eeq
where each piece is given by the first terms in  \eqref{usp8togl5}.

In particular, one can identify in the expansion \eqref{j expansion} all possible spinor bilinears with non-zero charge under $\xi$\footnote{Our notation is $\chi^T \gamma \chi'= \chi_{\alpha} \gamma^{\alpha \beta} \chi_{\beta}'$ and $\chi^{\dagger} \gamma \chi'= \chi^{* \alpha} \gamma_{\alpha}^{\, \, \beta} \chi_{\beta}'$ for a Cliff(5) element $\gamma$ and two $Spin(5)$ spinors $\chi$ and $\chi'$.}
\begin{eqnarray}\label{charged bilinears}
\jc_+^{m6}&=&4 \chi_1^T \gamma^{m} \chi_2 \nn \\
\jc_+^{mn}&=&-2(\chi_1^T \gamma^{mn} \chi_1 + \chi_2^T \gamma^{mn} \chi_2)  \nn \\
\jc_+^{mn6}&=&-2(\chi_1^T \gamma^{mn} \chi_1 - \chi_2^T \gamma^{mn} \chi_2)   \\
\jc_+^{mnp}&=&-4 \chi_1^T \gamma^{mnp} \chi_2  \nn \\
\jc_+^{7}&=&4 i \chi_1^T  \chi_2  \nn 
\end{eqnarray}
where we have defined
\beq\label{plusminusdef}
\jc_{\pm}=\jc_1 \pm i \jc_2 \ .
\eeq
The components of $\jc_-$ have exactly the same form with the replacement $\chi_i \to \chi_i^c$ and an overall minus sign in the above expressions.\footnote{For example, we have $\jc_-^{m6}=-4 \chi_1^{cT} \gamma^{m} \chi_2^c $.} On the other hand, $\jc_3$ is neutral since it is constructed from two oppositely charged spinors ($\chi$ and $\chi^{\dagger}$). The explicit expressions for the related bilinears are
\begin{eqnarray}\label{neutral3 bilinears}
\jc_3^{m6}&=&2(-\chi_1^{\dagger}\gamma^{m}\chi_2+\chi_2^{\dagger}\gamma^{m}\chi_1) \nn \\
\jc_3^{mn}&=&2(\chi_1^{\dagger} \gamma^{mn} \chi_1 + \chi_2^{\dagger} \gamma^{mn} \chi_2) \nn \\
\jc_3^{mn6}&=&2(\chi_1^{\dagger} \gamma^{mn} \chi_1 - \chi_2^{\dagger} \gamma^{mn} \chi_2)  \\
\jc_3^{mnp}&=&2(\chi_1^{\dagger}\gamma^{mnp}\chi_2+\chi_2^{\dagger}\gamma^{mnp}\chi_1) \nn \\
\jc_3^{7}&=&2i(-\chi_1^{\dagger}  \chi_2 + \chi_2^{\dagger}  \chi_1) \nn 
\end{eqnarray}
Together with those coming from $\kc$ \eqref{neutral0 bilinears}, these form the set of spinor bilinears which are neutral under the Killing vector $\xi$. Moreover, note that expansions similar to \eqref{k expansion} and \eqref{j expansion} can be done for the rescaled bispinors $\kcr$ and $\jcr$.

\subsection{Proof of the generalized integrability conditions}
\label{proof}

In this section we describe the general methodology used to prove the generalized integrability conditions \eqref{Hint}-\eqref{compint} from the Killing spinor equations \eqref{egspinor}-\eqref{dspinor}, while we relegate the details to the appendices. 

\subsubsection{Killing spinor equations}

In order to use the supersymmetry conditions efficiently, we need to turn the Killing spinor equations \eqref{egspinor}-\eqref{dspinor} into equations on $\jc_a$ and $\jc_0$. This can be done easily by taking the complex conjugate and transpose of the former. From the equation coming from requiring that the variation of the external gravitino equation vanishes, Eq. \eqref{egspinor}, we get

\noindent \textbf{External gravitino}

\begin{subequations}\label{susye}
\beq\label{susye1}
m\jc_{\pm}=\pm\jc_{\pm} G^E
\eeq
\beq\label{susye2}
m\jc_3=-\jc_0  G^E
\eeq
\beq\label{susye3}
m\jc_0=-\jc_3  G^E
\eeq
where
\beq
G^E=e^A\left[
 (\slashed{\partial}A) \Gamma^6 \CHS +i \dfrac{e^{\phi}}{4} \left((\slashed{F}_1 +\slashed{F}_5) \Gamma^6 - \slashed{F}_3\right)
  \right]
\eeq
\end{subequations}

\noindent From the requirement that the variation of the internal component of the gravitino vanishes, Eq. \eqref{igspinor}, we get

\noindent \textbf{Internal gravitino}

\begin{subequations}\label{derj}
\beq\label{derj0}
\nabla_m \jc_a =[\jc_a,G^{IS}_m]+\{\jc_a,G^{IA}_m\} ,\quad a=0,1,2,3
\eeq
where
\beq\label{derj symmetric}
G^{IS}_m=-\cl{4} \slashed{H}_m \Gamma^6 + \dfrac{ie^{\phi}}{8}(F_{1,m} + \slashed{F}_{3,m} \Gamma^6) \CHS - \dfrac{e^{\phi}}{8} (\ast F_5) \Gamma_{m6}
\eeq
\beq\label{derj antisymmetric}
G^{IA}_m=\dfrac{ie^{\phi}}{8} (F^p \Gamma_{mp}+\dfrac{F^{npq}}{3!} \Gamma_{mnpq} \Gamma^6) \CHS
\eeq
\end{subequations}

\noindent From requiring that the dilatino stays invariant, Eq. (\ref{dspinor}), we get

\noindent \textbf{Dilatino}

\begin{subequations}\label{susyd}
\beq\label{susyd1}
\jc_aG^D  =0 ,\quad a=0,1,2,3
\eeq
where
\beq
G^D =
\left[
  (\slashed{\partial} \phi) \Gamma^6 \Gamma_{(7)} - \dfrac{1}{2} \slashed{H} \CHS + \dfrac{ie^{\phi}}{2}\left(2 \slashed{F}_1 \Gamma^6 -\slashed{F}_3 \right)
  \right]
\eeq
\end{subequations}

\subsubsection{Integrability conditions}

Now, we are ready to prove the integrability conditions \eqref{Hint}-\eqref{compint} for the H and V structures. 
These are given in terms of the dressed objects $J_a, K$, but it turns out to be more tractable to work with the undressed objects $\jc$, $\kc$, in particular since the gauge fields and the derivative satisfy 
\beq
\tilde \mu \, D \equiv (\mu + \frac{2 \phi}{3}) D = 0 
\eeq
where $\tilde \mu$ is an element of $\mathfrak{e}_{6(6)} \oplus \mathbb R^+$. The dilaton appears here due to the way  it embeds in the $GL(5)$ piece in the adjoint action (see \eqref{embedgaugemn}, \eqref{embedgauge66}), and it reflects the fact that the (anti) fundamental representation is actually charged under the $\mathbb{R}^+$, i.e. we are working with objects which are dressed under the trombone (see \eqref{embedder} and \eqref{V and H rescaling}).

We will also use a crucial trick: the generalized integrability conditions stem from the generalized Lie derivative operation (\ref{GL}), which is independent of the generalized connection, as long as it is torsion free \cite{Coimbra:2011ky}. Thus, instead of embedding the partial derivative into the generalized derivative as in \eqref{embedder}, we are going to embed the covariant derivative, namely we will use as generalized connection the ordinary Levi-Civita connection. We thus have
\beq\label{embedder2}
D_{m6}= e^{2 \phi/3} \nabla_m \ .
\eeq

\noindent \textbf{$J_a$ equations}
\vspace*{.2cm}

Let us start with the moment map condition for the hyper-multiplet structure, Eq. \eqref{Hint}, that we repeat here 
\beq\label{Hint2}
D \tilde{J}_a + \kappa \,  \epsilon_{abc}
\tr( \tilde{J}_b, D\tilde{J}_c )=
\lambda_a   c(\tilde{K},\tilde{K}, \cdot)
\eeq
When undressing $J_a$, each term on the left hand side contributes two terms, one where the derivative is acting on the naked $\jc$, and another one with the derivative acting on $\mu$. Acting on the whole equation by 
$e^{-\mu}$ to undress it, we get the twisted moment map densities $\mm_a$
\begin{eqnarray} \label{twistmom}
\mm_a \equiv e^{ -\mu}\left(  D \tilde{J}_a + \kappa \,  \epsilon_{abc}
\tr( \tilde{J}_b, D\tilde{J}_c ) \right) =  \nn  \\
D \tilde{\jc_a} 
+\alg{D \mu}{\jcr_a}  
+\kappa \epsilon_{abc}  \tr\alg{\jcr_b}{D \jcr_c}
+\kappa \epsilon_{abc} \tr\alg{\jcr_b}{\alg{D \mu}{\jcr_c}}
\end{eqnarray}
where in analogy with their twisted counterparts \eqref{V and H rescaling}, we have defined the rescaled bispinors
\beq\label{jk calligraphic rescaling}
\tilde{\jc_a} = e^{2A-2\phi} \jc_a \ , \qquad
\tilde{\kc} = e^{-2\phi/3} \kc \ . 
\eeq

%
%
We are going to perform this calculation in \usp basis, where the derivative $D$ has components (cf. \eqref{e6f2})
\beq \label{DV}
D^{\alpha \beta}=  \dfrac{i e^{2\phi/3}}{2\sqrt{2}} (\Gamma^{m67})^{\alpha \beta} \nabla_m \equiv (\D^m)^{\alpha \beta} \nabla_m \ 
\eeq
where for later use we have defined the generalized vector $\D$, which has only a vectorial component along direction of the generalized derivative. We then get that \eqref{twistmom} reads, in \usp basis 
\beq\label{twisted moment map density}
M_a=[\nm \jcr_a, \D^m] 
+ (\alg{\nm \mu}{\jcr_a}  \D^m)
+\tr [\jcr_a  G^{IS}_m] \D^m
-\tr [(\nm \mu) \jcr_a)]\D^m \ .
\eeq
Here we have used the fact that the $\jc_a$ contain only a \rep{36} component (and thus the Killing form \eqref{killusp8} just reduces to a matrix trace)
and  in the third and fourth terms we have used the $\mathfrak{su}(2)$ algebra \eqref{su2algebra}. For the third term we also used the internal gravitino equation \eqref{derj}. The commutators $[ \ , \ ]$ and the traces are now understood as matrix commutators and traces respectively  ($v^m\propto \Gamma^{m67}$). The second term means the action of the adjoint element $\alg{\nm \mu}{\jcr_a} $ on the fundamental $v^m$.

Although \eqref{twisted moment map density}  seems not to be gauge-invariant ($\mu$ contains the gauge fields), this is not the case since the second and the fourth term together project onto the exterior derivative of the gauge fields, i.e. the fluxes. Using the internal and external gravitino equations \eqref{derj} and \eqref{susye} as well as the dilatino equations \eqref{susyd}, we find (see appendix \ref{The moment map for J} for the details of  this computation) 
\begin{subequations}\label{twistmom computed}
\beq\label{twistmom computed pm }
\mm_\pm = 0
\eeq
\beq\label{twistmom computed 3 }
\mm_3 = (-2 i m) \rho e^{-4 \phi/3}  \kc
\eeq
\end{subequations}
We thus verify the $\pm$ components of the moment map equations \eqref{Hint2}, for the choice $\lambda_\pm=0$, in agreement with \eqref{lambda values}. The third component $\mm_3$, should be, according to \eqref{Hint2} and \eqref{lambda values} proportional to the dual vector of $K$ through the cubic invariant. Indeed, one can check using the explicit form of ${\kc}$ in terms of spinors \eqref{K}, as well as the spinor normalizations \eqref{su2algebra extend} and the definition of the rescaled $\kc$ \eqref{jk calligraphic rescaling} that 
\beq
\big[ c(\tilde{\kc},\tilde{\kc}, \cdot) \big]^{\alpha \beta} = \rho e^{- 4\phi/3} \kc^{\alpha \beta} \ .
\eeq
We therefore verify the third component of the moment map equation with $\lambda_3 = - 2 i m$, in accordance to \eqref{lambda values}.

\vspace*{.3cm}

\noindent \textbf{$K$ and compatibility equations }
\vspace*{.2cm}

We rewrite here the integrability condition for  $K$ and the condition coming from requiring compatibility of the integrable H and V structures, Eqs \eqref{Vint} and \eqref{compint} 
\beq\label{Vint2}
\mathbb{{L}}_{\tilde{K}} \tilde{K} = 0
\eeq
\beq\label{compint2}
\mathbb{{L}}_{\tilde{K}}\tilde{J}_a = \dfrac{3i}{2} \epsilon_{abc} \lambda_b \tilde{J}_c \ .
\eeq
They both contain the Dorfman derivative along the (rescaled) twisted generalized vector $\tilde K=e^{-2\phi/3} K=e^{-2\phi/3} (e^{\mu} \kc)$. As before, it is convenient to split the contributions coming from the derivative of $\mu$ from the rest. Using the expression for the Dorfman derivative \eqref{GL}, one gets
\beq\label{twistd}
e^{-\mu} {\mathbb{L}}_{\tilde K}  = 
(\kcr \cdot \D) \Big(\nabla 
+ \nabla \mu  \Big)
-\Big( \D \times (\nabla \kcr  + (\nabla \mu) \kcr  ) \Big) 
\eeq
where the generalized vector $\D$ along the direction of the derivative $D$ was defined in \eqref{DV}. The first and third term are the same as in 
${\mathbb{L}}_{\kcr}$, while with the second and the fourth we define a twisted Dorfman derivative $\widehat{{\mathbb{L}}}_{\kcr}$, namely
\beq\label{twistdorfe6}
\widehat{\mathbb{L}}_{\kcr}\equiv e^{-\mu} {\mathbb{L}}_{\tilde K}= {\mathbb{L}}_{\kcr} +   (\kcr \cdot \D) \nabla \mu  
-  \, \D \times \big((\nabla \mu)\kcr \big) \,    \ .  
\eeq
Using this twisted derivative, we can now rewrite the integrability conditions \eqref{Vint2} and \eqref{compint2}  as equations on the undressed structures $\kc$ and $\jc$ (or rather their rescaled versions $\kcr$ and $\tilde{\jc}$ defined in \eqref{jk calligraphic rescaling}) as follows 
\beq\label{twistVint}
\widehat{\mathbb{L}}_{\mathcal{\tilde{K}}}\tilde{\kc} = 0
\eeq
\beq\label{twistcompint}
\widehat{\mathbb{L}}_{\mathcal{\tilde{K}}} \tilde{\jc}_a = \dfrac{3i}{2} \epsilon_{abc} \lambda_b \tilde{\jc}_c
\eeq

These equations turn out to be very simple using the fact that the twisted Dorfman derivative along $\tilde {\cal K}$  on spinor bilinears actually reduces to the usual Lie derivative along the vector part of ${\cal K}$ \cite{Ashmore:2015joa}, namely the Killing vector $\xi$ defined in \eqref{Killingdef}
\beq \label{dorf=lie}
\widehat{\mathbb{L}}_{\kcr} =  \mathcal{L}_{\xi} 
\quad \quad \text{on bispinors} \ .
\eeq
Let us show briefly why this is so. The derivative acting on a generic element  can be split as in a differential operator, corresponding to the first term in \eqref{GL}, and the rest, which is an algebraic operator from the point of view of the element that it acts on:
\beq \label{splitA}
\widehat{\mathbb{L}}_{\tilde \kc} = (\tilde \kc \cdot \D) \nabla + \ac  
\eeq
The first piece reduces to the directional derivative along the  Killing vector $\xi$. 
For the algebraic part, we decompose the operator ${\cal A}$, which acts in the adjoint, into the \usp pieces
\beq\label{splitA algebraic}
\ac= \ac|_{36} + \ac|_{42}
\eeq
and we have furthermore that $\ac|_{36}$ can be viewed as an element of $\text{Cliff(6)}$.  We show in the appendix \ref{The Dorfman derivative along K} that supersymmetry implies that 
\beq\label{dorf=lie algebraic}
\ac|_{36} = \cl{4} (\nabla_m \xi_n) \Gamma^{mn} , \qquad \ac|_{42}=0 
\eeq
Now let us consider the action of $\widehat{\mathbb{L}}_{\kc}$ on $\kc$ and $\jc_a$. These are respectively in the ${\bf 27}$ and ${\bf 36}$ of \usp, and combined they form the $\Rep{63}$, the representation of hermitean traceless bispinors, and thus we have simply
\beq\label{same cliff action}
\ac \, \kc = \cl{4} (\nabla_m \xi_n) [\Gamma^{mn}, \kc] , \qquad
\alg{\ac}{\jc_a} = \cl{4}(\nabla_m \xi_n) [\Gamma^{mn}, \jc_a]
\eeq
where the commutators are just gamma matrix commutators. Together with the directional derivative along $\xi$ from the first term in (\ref{splitA}), we conclude that $\widehat{\mathbb{L}}_{\kcr} =  \mathcal{L}_{\xi}$.

Using this, it is very easy to show \eqref{twistVint} and \eqref{twistcompint}. Given that the $Spin(5)$ spinors have a definite charge under this action, Eq. \eqref{spinorcharge}, the \usp spinors $\theta_{1,2}$ have charges $\pm (3im/2)$ and therefore the bispinors satisfy 
\beq
\mathcal{L}_{\xi} \mathcal{J}_{\pm} = \pm 3im \mathcal{J}_{\pm} \quad \text{and} \quad \mathcal{L}_{\xi} \mathcal{J}_3 = \mathcal{L}_{\xi} \mathcal{K} = 0 
\eeq
from which one can immediately verify \eqref{twistVint} and \eqref{twistcompint}.

Before closing this section, let us note that the fact that the twisted generalized Lie derivative along $\tilde \kc$ reduces to an ordinary Lie derivative along its vector part is actually a generic feature of ``generalized Killing vectors"\footnote{We thank C. Strickland-Constable for sharing this with us.}: it can be shown that if a generalized vector is such that the generalized Lie derivative along that vector on the objects defining the background --generalized metric for a generic background, and spinors or spinor bilinears for a supersymmetric one-- vanishes, then the Dorfman derivative along such a generalized vector reduces to an ordinary Lie derivative along its vector component \cite{Charlie}.

\section{The M-theory analogue}
\label{4}

In this section, we prove the generalized integrability conditions for compactifications of eleven-dimensional supergravity down to $AdS_5$. The situation is similar to the type IIB case since the group of global symmetries remains the same, namely \sx. However, the proof is more transparent since M-theory combines the degrees of freedom in a more compact form, avoiding thus the complications due to the $GL(5) \subset SL(6)$ embedding. In particular, the generalized tangent bundle is decomposed in this case as
\beq\label{e6 tangent bundle m-theory}
E\simeq T\mc 
\oplus \wedge^2 T^*\mc
\oplus \wedge^5 T^*\mc 
\eeq
where the internal manifold $\mc$ is now six-dimensional and the various terms correspond to momenta, M2- and M5-brane charges respectively. The latter can be dualized to a vector, and together with the first piece they form the $(\Rep{6}, \Rep{2})$ piece in the split of the fundamental ${\bf 27}$ representation under  $SL(6) \times SL(2)$ given in \eqref{funddecomp}. The derivative is embedded in one of the two components of this doublet appearing in the anti-fundamental representation\footnote{Note that here $D$ does not carry a rescaling factor in contrast to the type IIB case.}
\beq\label{embedder m-theory}
D^2_a=\nabla_a, \qquad D^1_a=D_{ab}=0  \ .
\eeq
%
%
The decomposition of the adjoint representation is given in \eqref{adjdecomp}, and the three-form gauge field $\mt$ embeds in $\mu$ as
\begin{subequations}\label{embedgauge m-theory}
\beq
\mu^1_{abc} = - (\star \mt)_{abc}
\eeq
\beq
\mu^2_{abc} = \mu^i_{\,j} =\mu^a_{\, \, b} =0 .
\eeq
\end{subequations}

The rescaled structures for M-theory are
\beq\label{V and H rescaling M}
\tilde{K} =  K \ , \qquad
\tilde{J}_a = e^{2A} J_a \ ,
\eeq
having the same form as for type IIB but with a vanishing dilaton.

Equations \eqref{Hint} and \eqref{Vint}  have exactly the same form as in the type IIB case, with  
\beq
\kappa= - \dfrac{i}{4\rt} e^{-3A}=(8i \rho e^{2A})^{-1}
\eeq
while \eqref{compint} has a different sign in our conventions, i.e.
\beq\label{compint m theory}
\mathbb{{L}}_{\tilde{K}}\tilde{J}_a = - \dfrac{3i}{2} \epsilon_{abc} \lambda_b \tilde{J}_c
\eeq
where again $\lambda_1 =\lambda_2 =0, \lambda_3 = -2im$. This sign difference is due to the fact the internal spinor has opposite charge compared to the type IIB case (cf. \eqref{spinor charge m theory}).

The supersymmetry variation of the gravitino (up to quadratic terms) reads\footnote{We use tildes for the eleven-dimensional gamma-matrices (see appendix \ref{Spinor conventions}).} 

\beq\label{11D susy variation}
\delta \Psi_M =
\nabla_M \epsilon
+\dfrac{1}{288} \Big(
\tilde{\Gamma}_M^{\,\,\,\,NPQR}-8\delta_M^N \tilde{\Gamma}^{PQR}
\Big)
\mtf_{NPQR} \epsilon
\eeq
where $G=d\mt$ and $\epsilon$ is the eleven-dimensional (Majorana) supersymmetry parameter.

The eleven-dimensional metric is written again in the form \eqref{metric ansatz} where now the internal metric $g_{ab}$\footnote{We use $a,b,c, \dots$ to describe representations of the $GL(6)$ group of diffeomorphisms of the internal manifold. Moreover, we will suppress from now on the $SU(2)_R$ adjoint index $a$ in $J_a$  in order to avoid confusion with the $GL(6)$ ones.} is six-dimensional and the spinor decomposition ansatz for M-theory compactifications reads
\beq\label{spinor ansatz m-theory}
\epsilon = \psi \otimes \theta + \psi^c \otimes \theta^c
\eeq
where $\theta$ is a complex 8-component spinor on the internal manifold. Finally, the field strength $\mtf$ is allowed to have only internal components in order to respect the isometries of $AdS_5$.

There is again a vector field $\xi$ which generates a symmetry of the full bosonic sector
\beq\label{Killingfull m theory}
\mathcal{L}_{\xi} \{g, A, G\}=0,
\eeq
where $\xi$ is now given by
\beq\label{Killing definition m theory}
\xi^a= \dfrac{i}{\rt} \theta^{\dagger} \Gamma^{a7} \theta
\eeq

One can construct the H and V structures in exactly the same way as for the type IIB case. In particular, the expressions \eqref{normusp8} to \eqref{rho} have exactly the same form where 
\beq
\theta_1 = \theta \ , \qquad \theta_2 = \theta^c  \ . 
\eeq
However, the $\theta_i$ are not constructed from two $Spin(5)$ spinors as in type IIB. 

The decomposition of the supersymmetry variation \eqref{11D susy variation} in external and internal pieces is similar to the type IIB case with the difference that here we do not have a dilatino variation.  In terms of $\jc_a$ and $\jc_0$, we get the differential condition
\beq\label{derj m-theory}
\nabla_a \jc =[\jc ,G^{IS}_a]+\{\jc,G^{IA}_a\} ,\quad \jc=\jc_{\pm}, \jc_3, \jc_0
\eeq
where 
\beq\label{derj m-theory sym-asym}
G^{IS}_a=-\cl{36} \mtf_{abcd} \Gamma^{bcd}, \qquad
G^{IA}_a=-\dfrac{i}{12} (\star \mtf)_{ab} \Gamma^{b7}
\eeq
and the algebraic ones
\beq\label{susye1 m-theory}
m\jc_{\pm}=\pm\jc_{\pm}G^E
\eeq
\beq\label{susye2 m-theory}
m\jc_3=-\jc_0 G^E
\eeq
\beq\label{susye3 m-theory}
m\jc_0=-\jc_3 G^E
\eeq
where now $G^E$ is given by
\beq\label{susye4 m-theory}
G^E=e^A\left[
 (\slashed{\partial}A) \Gamma^7  + \dfrac{i}{12} (\star \mtf)_{ab} \Gamma^{ab}
  \right]
\eeq

The Clifford expansion for $\kc$ is now
\beq\label{k expansion m theory}
\kc=\dfrac{1}{2\rt}\Big[\zeta_a \Gamma^a +i \xi_a \Gamma^{a7} 
+\dfrac{i}{2} V_{ab} \Gamma^{ab7} \Big]
\eeq
where the components correspond to the different pieces in the $SL(6)$ decomposition of the fundamental, Eq. \eqref{e6 tangent bundle m-theory}, 
and correspond to the following spinor bilinears
\beq\label{neutral0 bilinears m theory}
\zeta^{a}=\cl{\rt} \theta^{\dagger} \Gamma^{a} \theta, \qquad
V^{ab}=\dfrac{i}{\rt}  \theta^{\dagger} \Gamma^{ab7} \theta
\eeq
and the vector $\xi$ is the Killing vector defined in \eqref{Killing definition m theory}.

For the triplet $\jc$, the expansion reads
\beq\label{j expansion m theory}
\jc =-\dfrac{1}{8}\Big[\cl{2}\jc^{ab} \Gamma_{ab}
- \jc^7 \Gamma_7
+\cl{6}\jc^{abc}  \Gamma_{abc}  \Big]
\eeq
where now the the analogue of the \eqref{adj36 split} split under $GL(6)$ is 
\beq\label{adj36 split m theory}
\Rep{36}=\Rep{15}+\Rep{1}+\Rep{20}
\eeq
The components of ${\cal J}_+$ are given by the following spinor bilinears, all charged under $\xi$ 
\beq\label{charged bilinears m theory}
\jc_+^{ab}=-2 \theta^T \Gamma^{ab} \theta ,\qquad
\jc_+^{abc}=-2 \theta^T \Gamma^{abc} \theta, \qquad
\jc_+^{7}=-2 \theta^T \Gamma^{7} \theta
\eeq
and the corresponding expressions for $\jc_-$ are given by the replacement $\theta \to \theta^c$ and an overall minus sign. For $\jc_3$, the analogous expressions are
\beq\label{neutral3 bilinears m theory}
\jc_3^{ab}=2 \theta^{\dagger} \Gamma^{ab} \theta, \qquad
\jc_3^{abc}=2 \theta^{\dagger} \Gamma^{abc} \theta, \qquad
\jc_3^{7}=2 \theta^{\dagger} \Gamma^{7} \theta
\eeq

The procedure to prove the integrability conditions is the same as the one described in subsection \ref{proof} for type IIB. In particular, we again work with the undressed structures $\kc$ and $\jc$ and with the twisted moment map density and the twisted Dorfman derivative defined in \eqref{twistmom} and \eqref{twistd} respectively for type IIB. We leave the details of this calculation to the appendices. The key point that the twisted Dorfman derivative along $\kc$ reduces to the ordinary Lie derivative along $\xi$, Eq. \eqref{dorf=lie}, is also true here and from \eqref{spinor charge m theory}, we get
\beq
\mathcal{L}_{\xi} \mathcal{J}_{\pm} = \mp 3im \mathcal{J}_{\pm} \quad \text{and} \quad \mathcal{L}_{\xi} \mathcal{J}_3 = \mathcal{L}_{\xi} \mathcal{K} = 0 
\eeq

\section{Discussion}
\label{sec:Discussion}

We have proven that the supersymmetry equations relevant for $AdS_5$ vacua with generic fluxes preserving eight supercharges in type IIB and M-theory compactifications translate into the integrability conditions  \eqref{Hint},\eqref{Vint} and \eqref{compint} in Exceptional Generalized Geometry. The integrability conditions involve generalized structures in the fundamental and adjoint representations of the $E_{6(6)}$ U-duality group. Although our calculations were performed for the particular case of $AdS_5$ compactifications, the integrability conditions are expected to be the same for other $AdS_d$ vacua of type II (either IIA or IIB) and M-theory compactifications preserving eight supercharges, since these are described by vector and hypermultiplets. 
A particularly interesting case to analyze is that of $AdS_4$ vacua, where the relevant U-duality group is $E_{7(7)}$, with maximal compact subgroup $SU(8)$. The construction of the generalized structures from spinor bilinears is the same, and since  our calculations were done in \usp language, the extension to $SU(8)$ should be rather straightforward.  

The description of $AdS_5$ vacua in exceptional generalized geometry has nice applications in AdS/CFT. The original example is the $AdS_5 \times S^5$ solution supported by five-form flux (in the type IIB case) which is dual to $\nc=4$ SYM. Allowing for generic internal manifolds (and fluxes) but still preserving some supersymmetry corresponds to supersymmetric deformations on the field theory side. $AdS$ vacua
are dual to deformations that preserve conformal invariance on the gauge theory. Having a compact description of the internal geometry opens then the way for finding the supergravity dual of these deformations in a rather systematic way, as very recently shown in \cite{Ash}. We will explore this direction further in future work.

\acknowledgments{We would like to thank Anthony Ashmore, Mirela Babalic, Michela Petrini, Carlos Shahbazi, Charlie Strickland-Constable, Orestis Vasilakis and Daniel Waldram for insightful discussions. This work was supported in part by the ERC Starting Grant 259133  {\em ObservableString}, and by the P2IO LabEx (ANR-10-LABX-0038) in the framework Investissements d'Avenir � (ANR-11-IDEX-0003-01) managed by Agence Nationale de la Recherche.}

\appendix

\section{Spinor conventions}
\label{Spinor conventions}

In the paper we use spinors of $Spin(1,4)$ and $Spin(5)$ and $Spin(1,9)$ for type IIB, and $Spin(6)$ and $Spin(1,10)$ in M-theory. We give our conventions for all of them, explain their relations and provide some useful formulae for our calculations. In this section, all the indices are meant to be flat.

For five Euclidean dimensions, the gamma matrices are denoted by $\gamma^m, \quad m=1,\dots5$ and satisfy\footnote{An explicit construction of them can be given by $\gamma^1 =\sigma^1 \otimes \sigma^0  ,
\gamma^2 =\sigma^2 \otimes \sigma^0  ,
\gamma^3 =\sigma^3 \otimes \sigma^1  ,
\gamma^4 =\sigma^3 \otimes \sigma^2$ and $ \gamma^5=-\sigma^3 \otimes \sigma^3  $, in which case the interwiner is $C_5=\sigma^1 \otimes \sigma^2$.}
\begin{subequations}
\begin{align}
(\gamma^m)^{\dagger}&=\gamma^m \\
(\gamma^m)^T &= C_5 \gamma^m C_5^{-1} \\
(\gamma^m)^* &= D_5 \gamma^m D_5^{-1} 
\end{align}
\end{subequations}
where we take $D_5=C_5$ and we have $\gamma^{12345}=1_4$. For a spinor $\chi$, the conjugate spinor is defined as
\beq
\chi^c= D_5^* \chi^*
\eeq
and satisfies the properties
\beq
(\gamma^m \chi)^c = \gamma^m \chi^c,\qquad D_5^* D_5 =-1 \Rightarrow \chi^{cc} =-\chi
\eeq

For the 5-dimensional external space, we have a Lorentzian version of the above. The intertwining relations for the gamma matrices $\rho^{\mu}$ are\footnote{Explicitly we can take $\rho^0 =i \sigma^2 \otimes \sigma^0,\rho^i =\sigma^1 \otimes \sigma^i$ with $i=1,2,3$, $\rho^4 =- \sigma^3 \otimes \sigma^0 $ and $A_{1,4}= \rho^0, C_{1,4}=\rho^0 \rho^2$.}
\begin{subequations}
\begin{align}
(\rho^{\mu})^{\dagger}&=-A_{1,4} \rho^{\mu} A^{-1}_{1,4}  \\
(\rho^{\mu})^T &= C_{1,4} \rho^{\mu} C_{1,4}^{-1} \\
(\rho^{\mu})^* &= -D_{1,4} \rho^{\mu} D_{1,4}^{-1}
\end{align}
\end{subequations}
where $\mu=0,\dots,4$, $\rho^{01234} =-i \, 1_4 $ and $D_{1,4}=-C_{1,4} A_{1,4}$. The conjugate spinor is defined as
\beq
\psi^c= D_{1,4}^* \psi^*
\eeq
and satisfies
\beq
(\rho^{\mu} \psi)^c = - \rho^{\mu} \psi^c,\qquad D_{1,4}^* D_{1,4} =-1 \Rightarrow \psi^{cc} =-\psi
\eeq

Now, let us combine the above representations to construct a 10-dimensional Clifford algebra. We define
\begin{subequations}\label{10decompogamma}
\begin{align}
&\hat{\Gamma}^{\mu}=\rho^{\mu} \otimes 1_4 \otimes \sigma^3, \quad \mu=0,\dots,4 \\
&\hat{\Gamma}^{m+4}= 1_4 \otimes \gamma^m \otimes \sigma^1, \quad m=1,\dots,5
\end{align}
\end{subequations}
The last factor is needed to allow for a chirality matrix in 10 dimensions:
\begin{equation}
\hat{\Gamma}_{(11)}= \hat{\Gamma}^0\dots\hat{\Gamma}^9=1_4 \otimes 1_4 \otimes \sigma^2
\end{equation}
The 10-dimensional interwiners are constructed as follows
\begin{subequations}
\begin{align}
&A_{1,9}=-A_{1,4} \otimes 1_4 \otimes \sigma^3 \, \Longrightarrow \, (\hat{\Gamma}^M)^{\dagger}= -A_{1,9} \hat{\Gamma}^M A_{1,9}^{-1} \\
&C_{1,9}=C_{1,4} \otimes C_5 \otimes \sigma^2 \, \Longrightarrow \, (\hat{\Gamma}^M)^T= -C_{1,9} \hat{\Gamma}^M C_{1,9}^{-1} \\
&D_{1,9}=D_{1,4} \otimes D_5 \otimes \sigma^1 \,\Longrightarrow \, (\hat{\Gamma}^M)^*= D_{1,9} \hat{\Gamma}^M D_{1,9}^{-1}
\end{align}
\end{subequations}
A 10-dimensional spinor $\epsilon$ splits as
\beq
\epsilon = \psi \otimes \chi \otimes u
\eeq
where $u$ is acted upon by the Pauli matrices. For the conjugate spinor we have 
\beq
\epsilon^c= D_{1,9}^* \epsilon^*, \qquad D_{1,9}^* D_{1,9}=1\Rightarrow \epsilon^{cc}=\epsilon
\eeq
The type IIB Majorana-Weyl spinors $\epsilon_i$ are  
\beq\label{10decompospinorapp}
\epsilon_i=\psi \otimes \chi_i \otimes u + \psi^c \otimes \chi_i^c \otimes u, \quad i=1,2
\eeq
Their chirality and reality properties require
\beq\label{u properties}
u=\sigma^2 u = \sigma^1 u^*
\eeq

We construct now gamma matrices $\Gamma^a , \, a=1,\dots6$ for Cliff(6)  from our representation for Cliff(5). We define
\begin{equation}
\Gamma^m=
\begin{pmatrix}
0 & \gamma^m \\
\gamma^m & 0
\end{pmatrix}, \quad
m=1,\dots,5, \qquad
\Gamma^6=
\begin{pmatrix}
 1 & 0 \\
 0 & -1
\end{pmatrix}
\end{equation}
\beq
\Gamma_7= i \Gamma^1 \dots \Gamma^6 = 
\begin{pmatrix}
0 & -i \\
i & 0
\end{pmatrix}, \qquad
i \Gamma^{67}=
\begin{pmatrix}
0 & 1 \\
1 & 0
\end{pmatrix}
\eeq
The interwiner for Cliff(6) is 
\beq \label{C6}
C=C^{\alpha \beta}=\dg{C_5}{C_5},\qquad C^{-1}=\dab{C}=\dg{C_5^{-1}}{C_5^{-1}}
\eeq
which raises and lowers spinor indices as  $\uab{\Gamma}=C^{\alpha \gamma} \Gamma_{\gamma}{}^{\beta} $,  $\dab{\Gamma}=\Gamma_{\alpha}{}^{\gamma}C_{\gamma \beta}$. For any Cliff(6) element $\Gamma$, we have
\beq
\Gamma^{(n)}_{\beta \alpha}=-(-)^{Int[n/2]}\dab{\Gamma^{(n)}}
\eeq
while the reality properties read\footnote{All the $C$'s defined in this section are antisymmetric, hermitian and unitary.}
\beq
\Gamma_a^*=C\Gamma_a\inv{C}
\eeq
The 6-dimensional gamma matrices act on USp(8) spinors $\theta_\alpha$, $\alpha=1,..8$. In the main text, we use the following 
\beq
\theta_1=\begin{pmatrix}
\chi_1 \\ \chi_2
\end{pmatrix},
\quad
\theta_2=\begin{pmatrix}
\chi_1^c \\ \chi_2^c
\end{pmatrix}
\eeq
satisfying
\beq
\theta^{*i \alpha}= (-i \sigma^2)^{ij} C^{\alpha \beta} \theta_{j \beta}
\eeq

The eleven-dimensional gamma-matrices relevant for M-theory can be built directly from the six-dimensional ones $\Gamma^a$ constructed above and  from the $\rho^{\mu}$ of $AdS_5$ as follows
\begin{subequations}\label{11decompogamma}
\begin{align}
&\tilde{\Gamma}^{\mu}=\rho^{\mu} \otimes \Gamma_7, \quad \mu=0,\dots,4 \\
&\tilde{\Gamma}^{a+4}= 1_4 \otimes \Gamma^a , \quad a=1,\dots,6
\end{align}
\end{subequations}
The relevant interwiners for eleven dimensions are
\begin{subequations}
\begin{align}
&C_{1,10}=C_{1,4} \otimes C_6 \Gamma_7 \, \Longrightarrow \, (\tilde{\Gamma}^M)^T= -C_{1,10} \tilde{\Gamma}^M C_{1,10}^{-1} \\
&D_{1,10}=D_{1,4} \otimes D_6 \,\Longrightarrow \, (\tilde{\Gamma}^M)^*= D_{1,10} \tilde{\Gamma}^M D_{1,10}^{-1}
\end{align}
\end{subequations}
A spinor in eleven dimensions $\epsilon$ decomposes
as
\beq
\epsilon = \psi \otimes \theta
\eeq
while the conjugate spinor is given by 
\beq
\epsilon^c = D^*_{1,10} \epsilon = \psi^c \otimes \theta^c
\eeq
The Majorana property of the M-theory supersymmetry parameter requires then
\beq
\epsilon = \psi \otimes \theta + \psi^c \otimes \theta^c
\eeq

We finish by giving some Fierz identities which are heavily used in our calculations

\beq\label{fierz 4}
(\Gamma_{ab7})^{\alpha \beta} (\Gamma^{ab7})_{\gamma \delta} -2 (\Gamma_a )^{\alpha \beta} (\Gamma^a)_{\gamma \delta}
+2 (\Gamma_{a7} )^{\alpha \beta} (\Gamma^{a7})_{\gamma \delta} 
= 16 \delta^{\alpha}_{[\gamma}\delta^{\beta}_{\delta]} + 2 C^{\alpha \beta} C_{\gamma \delta}
\eeq
\beq\label{fierz 1}
\asym{\Gamma^{(a}}{\Gamma^{b)7}}= \cl{6} g^{ab} \asym{\Gamma^{c}}{\Gamma^{c7}}
\eeq
\beq\label{fierz 2}
\asym{\Gamma^{[a}}{\Gamma^{b]7}}=
-\asym{\Gamma^{ab7}}{C} =
- \dfrac{i}{24}\epsilon_{abcdef} \asym{\Gamma^{cd7}}{\Gamma^{ef7}}
\eeq
\beq\label{fierz 3}
\asym{\Gamma^{[a|7}}{\Gamma^{bc]7}}=
\asym{\Gamma^{a7}}{\Gamma^{bc7}}+
2g^{a[b} \asym{\Gamma^{c]}}{C}
\eeq
\beq\label{fierz 6}
\Gamma^{6}_{[\alpha \beta}   \Gamma^{m}_{\gamma \delta]}  = -\Gamma^{67}_{[\alpha \beta}  \Gamma^{m7}_{\gamma \delta]}
\eeq
\beq\label{fierz 7}
\asym{\Gamma^{m67}}{\Gamma^{np7}}+\asym{\Gamma^{mnp6}}{C}=
2g^{m[n} \asym{\Gamma^{p]}}{\Gamma^6}
\eeq
\beq\label{fierz 5}
\gamma_{mn}^{\alpha \beta} (\gamma^{mn})^{\gamma \delta} = 
10 C_5^{\alpha \beta} C_5^{\gamma \delta}
+6 \gamma_{m}^{\alpha \beta} (\gamma^{m})^{\gamma \delta}
+8 \gamma_{m}^{\alpha \gamma} (\gamma^{m})^{\beta\delta}
\eeq
Let us note that one can derive additional Fierz identities by exploiting the following Leibniz-like rule:
\beq\label{leibniz-like}
\asym{A}{B}=\asym{C}{D} \Longrightarrow 
\asym{(A\Gamma)}{B} + \asym{A}{(B\Gamma)} = 
\asym{(C\Gamma)}{D} + \asym{C}{(D \Gamma)}
\eeq
for any antisymmetric elements $A,B,C,D$ and $\Gamma$ of Cliff(6).

\section{$E_{6(6)}$ representation theory}\label{e6reps}

The group \sx is a particular real form of the $E_6$ family of Lie groups. It is generated by 78 elements, out of which 36 are compact and 42 are not. It contains as subgroups \usp and \sls.

\subsection{\sls decomposition}

The vector representation $V$ of \sx is 27-dimensional and splits under \sls as 
\begin{subequations}
\beq
\Rep{27}=(\overline{\Rep{6}}, \Rep{2}) + (\Rep{15}, \Rep{1}), \quad V=(V^{i}_{\; a}, V^{ab})
\eeq
while we will also need its dual
\beq
\overline{\Rep{27}}=(\Rep{6}, \overline{\Rep{2}}) + (\overline{\Rep{15}}, \Rep{1}), \quad Z=(Z^{a}_{\; i}, Z_{ab})
\eeq
\end{subequations}
The adjoint decomposes  
\beq
\Rep{78}=(\Rep{35}, \Rep{1}) + (\Rep{1}, \Rep{3}) + (\overline{\Rep{20}}, \Rep{2}), \quad \mu = (\mu^a_{\; b},\mu^i_{\; j},  \mu^i_{abc})
\eeq
and its action on the vector is given by
 \begin{subequations}\label{e6v}
\beq\label{e6v1}
\left(\mu \,  V\right)^{i}_{\; a} = - \mu^b_{\; a} V^{i}_{\; b} + \mu^i_{\; j} V^{j}_{\; a}
+\dfrac{1}{2} \mu^i_{abc} V^{bc}
\eeq
\beq\label{e6v2}
\left(\mu \,  V\right)^{ab} =  \mu^a_{\; c} V^{cb} - \mu^b_{\; c} V^{ca} 
- \epsilon_{ij} (\star \mu^i)^{abc} V^{j}_{\; c}
\eeq
\end{subequations}
while on the dual vector by
\begin{subequations}\label{e6dv}
\beq\label{e6dv1}
(\mu \,  Z)_i^a = \mu^a_{\; b} Z_i^b - \mu^j_{\; i} Z_j^a - \dfrac{1}{2} \epsilon_{ij} (\star \mu^j)^{abc} Z_{bc}
\eeq
\beq\label{e6dv2}
(\mu \, Z)_{ab}=  - \mu^c_{\; a} Z_{cb} + \mu^c_{\; b} Z_{ca} - \mu^i_{abc} Z_i^c
\eeq
\end{subequations}
where $a,b,c,\dots$ run from 1 to 6 and $i,j$ from 1 to 2.

The $\mathfrak{e}_{6(6)}$ algebra $\alg{\mu}{\nu}$ is
\begin{subequations}\label{e6alg}
\beq\label{e6alg1}
\alg{\mu}{\nu}^i_{\,j}=\mu^i_{\,k} \nu^k_{\,j} +\cl{12} \mu^i_{abc} \epsilon_{jk} (\star \nu^k)^{abc} - (\mu \leftrightarrow \nu)
\eeq
\beq\label{e6alg2}
\alg{\mu}{\nu}^a_{\,b}=\mu^a_{\,c} \nu^c_{\,b} -\dfrac{1}{4} \mu^i_{bcd} \epsilon_{ij} (\star \nu^j)^{acd} - (\mu \leftrightarrow \nu)
\eeq
\beq\label{e6alg3}
\alg{\mu}{\nu}^i_{abc}=\mu^i_{\,j} \nu^j_{abc} -3 \mu^d_{\,[a} \nu^i_{bc]d} - (\mu \leftrightarrow \nu)
\eeq
\end{subequations}

The group \sx  has a quadratic and a cubic invariant. Given a vector $V$ and a dual vector $Z$, the quadratic invariant is
\beq\label{e6binv}
b(V,Z) = V^{i}_{\; a} Z_i^a + \dfrac{1}{2} V^{ab} Z_{ab}
\eeq
while the cubic is given by
\beq\label{e6cinv}
c(V,U,W)=\cl{2\rt}  \epsilon_{ij} \bigg(
V^{ab}U^i_a W^j_b +U^{ab}V^i_a W^j_b +  W^{ab}V^i_a U^j_b 
\bigg)
-\cl{16 \rt} \epsilon_{abcdef}V^{ab}U^{cd} W^{ef}
\eeq
where $U,V$ and $W$ are all in the fundamental. 
This allows to construct a dual vector from two vectors by ``deleting'' one of the vectors in the cubic invariant, namely
\begin{subequations}\label{e6vvtodv}
\beq\label{e6vvtodv1}
[c(V,U, \cdot)]_i^a =\cl{2\rt}  \epsilon_{ij}\left(V^{ab}{U}^{j}_{b} + {U}^{ab}V^j_b \right) 
\eeq
\beq\label{e6vvtodv2}
[c(V,U, \cdot)]_{ab}= \dfrac{1}{\sqrt{2}}\epsilon_{ij} V^i_{[a} {U}^j_{b]}
-\cl{8 \rt}  \epsilon_{abcdef}V^{cd} {U}^{ef}
\eeq
\end{subequations}


\subsection{\usp decomposition}

The other subgroup of $E_{6(6)}$ that we use is \usp. The \rep{27} fundamental representations of $E_{6(6)}$ is irreducible under \usp, and encoded  by an antisymmetric traceless tensor
\beq
V=V^{\alpha \beta} 
\eeq
with $V^\alpha{}_\alpha=0$. The \usp indices $\alpha , \beta , \dots$ are raised and lowered with $\dab{C}$ in \eqref{C6} , which plays the role of \usp symplectic invariant.

The adjoint decomposes as 
\beq\label{usp8 adjoint splitting}
\Rep{78} = \Rep{36} + \Rep{42}, \quad \mu = (\mu^{\alpha}_{\;\beta}, \mu^{\alpha \beta \gamma \delta})
\eeq
with $\mu_{\alpha\beta}=\mu_{\beta\alpha}$, $\mu^{\alpha\beta\gamma\delta}=\mu^{[\alpha\beta\gamma\delta]}$ and $\mu^{\alpha \beta \gamma \delta}C_{\gamma \delta}=0$. 
Furthermore, in our conventions we have 
\beq
\mu^*_{\alpha\beta} = - \mu_{\alpha\beta}, \qquad
\mu^*_{\alpha\beta\gamma\delta}=\mu_{\alpha\beta\gamma\delta}
\eeq
The adjoint action is
\beq\label{usp8actiona}
(\mu  V)^{\alpha \beta}= \mu^{\alpha}_{\;\gamma} V^{\gamma \beta} - \mu^{\beta}_{\;\gamma} V^{\gamma \alpha} - \mu^{\alpha \beta \gamma \delta} V_{\gamma \delta}
\eeq
\beq\label{usp8actionb}
(\mu  Z)^{\alpha \beta}= \mu^{\alpha}_{\;\gamma} Z^{\gamma \beta} - \mu^{\beta}_{\;\gamma} Z^{\gamma \alpha} + \mu^{\alpha \beta \gamma \delta} Z_{\gamma \delta}
\eeq
and the  $\mathfrak{e}_{6(6)}$ algebra is given by 
\begin{subequations}\label{e6algusp8}
\beq\label{e6alg1usp8}
\alg{\mu}{\nu}_{\alpha \beta}= 
\mu_{\alpha}^{\, \,\gamma} \nu_{\gamma \beta}
-\cl{3} \mu_{\alpha}^{\, \, \gamma \delta \epsilon} \nu_{\gamma \delta \epsilon \beta}
- (\mu \leftrightarrow \nu) \ . 
\eeq
\beq\label{e6alg2usp8}
\alg{\mu}{\nu}_{\alpha \beta \gamma \delta}=
-4\mu_{[\alpha}^{\, \,\,\epsilon}  \nu_{\beta \gamma \delta ] \epsilon}
- (\mu \leftrightarrow \nu)
\eeq
\end{subequations}
The quadratic and the cubic invariant of \sx take a particularly simple form in the \usp basis
\beq
b(V,Z) = V^{\alpha\beta} Z_{\beta \alpha}
\eeq
and
\beq
c(V,U,W)= V^{\alpha}_{\,\,\,\beta} U^{\beta}_{\,\,\,\gamma} W^{\gamma}_{\,\,\,\alpha}
\eeq
and we also have 
\beq\label{dualvectsl6}
[c(V,U, \cdot)]^{\alpha \beta}=\cl{2}(V^{\alpha}_{\,\,\,\gamma} V'^{\gamma \beta} - V^{\beta}_{\,\,\,\gamma} V'^{\gamma \alpha}- \cl{4} C^{\alpha \beta} V^{\gamma \delta} V'_{\delta \gamma})
\eeq
In our calculations we also need the adjoint projection built out of a vector $V$ and a dual vector $Z$. This is given by
\begin{subequations}\label{adjprojusp8}
\beq\label{adjprojusp8a}
(V \times Z)^{\alpha \beta}= 2 V^{(\alpha}_{\quad \gamma} Z^{|\gamma| \beta)}
\eeq
\beq\label{adjprojusp8b}
(V \times Z)^{\alpha \beta \gamma \delta} = 6 \Big(
\brab{V} \cdbr{Z} 
+V^{[\alpha}_{\quad \epsilon} Z^{|\epsilon| \beta} \cdbr{C} 
+\cl{3} (V^{\epsilon}_{\,\,\, \zeta} Z^{\zeta}_{\,\,\, \epsilon} ) 
\brab{C} \cdbr{C} 
\Big)
\eeq
\end{subequations}
Finally, the Killing form is 
\beq\label{killusp8}
\tr (\mu, \nu) = \mu^{\alpha \beta} \nu_{\alpha \beta} 
+ \cl{6} \mu^{\alpha \beta \gamma \delta} \nu_{\alpha \beta \gamma \delta}
\eeq

\subsection{Transformation between \sls and \usp }

Our calculations involve objects which are more naturally described in the \sls basis (gauge fields and derivative) and others (spinors) which have a natural \usp description. Therefore, it is useful to have explicit formulae for the transformation rules between them. For this purpose, we use the gamma matrices $\Gamma^a$ defined in 6 dimensions. It's also useful to introduce two sets of them:
\beq
\Gamma^a_i=(\Gamma^a, i \Gamma^a \CHS), \quad \quad i=1,2
\eeq
The transformation rules for the vector (fundamental) and the dual vector (anti-fundamental) representation are
\begin{subequations}\label{e6f}
\beq\label{e6f1}
V^{\alpha \beta} =  \dfrac{1}{2\sqrt{2}} (\Gamma^a_i)^{\alpha \beta} V^{i}_{\; a} + \dfrac{i}{4\sqrt{2}} (\Gamma_{ab7} )^{\alpha \beta} V^{ab} 
\eeq
\beq\label{e6f2}
Z^{\alpha \beta} =  \dfrac{1}{2\sqrt{2}} (\Gamma^i_a)^{\alpha \beta} Z^{a}_{\; i} + \dfrac{i}{4\sqrt{2}} (\Gamma^{ab7})^{\alpha \beta} Z_{ab} 
\eeq
\end{subequations}
and are easily inverted
\begin{subequations}\label{e6inv}
\beq
V^{i}_{\; a}=\cl{2 \rt} V^{\alpha \beta} (\Gamma_{a}^i)_{\beta \alpha}, 
\quad \quad
V^{ab}=\dfrac{i}{2 \rt}V^{\alpha \beta}(\Gamma^{ab7} )_{\beta \alpha}
\eeq
\beq
Z^{a}_{\; i}=\cl{2\rt} Z^{\alpha \beta} (\Gamma_{i}^a)_{\beta \alpha}, 
\quad \quad
Z_{ab}=\dfrac{i}{2 \rt}Z^{\alpha \beta}(\Gamma_{ab7})_{\beta \alpha}
\eeq
\end{subequations}
For the adjoint representation we have\footnote{$SL(2)$ indices are raised and lowered with $\delta_{ij}$.}
\begin{subequations}\label{gl5tousp8}
\beq\label{gl5tousp8a}
\dab{\mu}=\dfrac{1}{4}\bigg[\mu^a_{\,\,b}\dab{(\Gamma_a^{\,\,b})}+i\epsilon_i^{\,\,j}\mu^i_{\,\,j} \dab{(\CHS)}
+\dfrac{1}{6} \epsilon_i^{\,\,j} \mu^i_{abc} \dab{(\Gamma^{ab} \Gamma^c_j)}\bigg]
\eeq 
\begin{align}\label{gl5tousp8b}
 \nonumber \mu^{\alpha\beta\gamma\delta}=\cl{8}\Bigg[
&- \mu^a_{\,\,b} \bigg( \brab{(\Gamma^i_a)} \cdbr{(\Gamma^b_i)} - \brab{(\Gamma_{ac7})} \cdbr{(\Gamma^{cb7})}\bigg) \\ \nonumber
&+ \mu^i_{\,\,j} \brab{(\Gamma^a_i)} \cdbr{(\Gamma^j_a)} \\ 
& +i \mu^i_{abc} 
\brab{(\Gamma^a_i)} \cdbr{(\Gamma^{bc7})}
\Bigg]
\end{align}
\end{subequations}
Their inverses are given by
\begin{subequations}\label{usp8togl5}
\beq\label{usp8togl5a}
\mu^a_{\,\,b}= 
-\cl{4} \uab{\mu} \dba{(\Gamma^a_{\,\,b})}
-\dfrac{1}{16} \mu^{\alpha \beta \gamma \delta}
\asym{(\Gamma_i^a)}{(\Gamma^i_b)}
\eeq
\beq\label{usp8togl5b}
\mu^i_{\,\,j}= 
-\dfrac{i}{4} \epsilon^i_{\,\,j} \uab{\mu} \dba{(\Gamma_7)}
+\dfrac{1}{48} \mu^{\alpha \beta \gamma \delta}
\asym{(\Gamma^i_a)}{(\Gamma^a_j)}
\eeq
\beq\label{usp8togl5c}
\mu^i_{abc} =
- \dfrac{i}{4} \uab{\mu} \dba{(\Gamma^i_a \Gamma_{bc7})}
+\dfrac{i}{8} \mu^{\alpha \beta \gamma \delta}
\asym{(\Gamma^i_a)}{(\Gamma_{bc7})}
\eeq
\end{subequations}

\section{Some constraints from supersymmetry}
\label{SUSY and the Killing vector}

In this section we are going to prove some useful conditions that the spinor bilinears in \eqref{charged bilinears}, \eqref{neutral3 bilinears}, \eqref{neutral0 bilinears}, \eqref{charged bilinears m theory}, \eqref{neutral3 bilinears m theory}, \eqref{neutral0 bilinears m theory} and \eqref{Killing definition m theory} satisfy and which serve as an intermediate step in order to derive the integrability conditions \eqref{Hint}-\eqref{compint}. The most important relations are also stated in the main text. We split into the bilinears in type IIB, and those of M-theory.

\subsection{Type IIB}
\label{app:xsi}

Let us start by studying the vector $\xi$ defined in \eqref{Killingdef}. By tracing \eqref{derj} with $\Gamma_{n67}$, we get
\beq\label{Killingproof}
\nm \xi_n = - \cl{2} \zeta_7^p H_{mnp} + \dfrac{\ef}{4} \zeta^p F_{mnp}
+\dfrac{\ef}{4} (*V)_{mnp} F^p + \dfrac{\ef}{4} V_{mn} (\ast F_5)
\eeq
Since the right hand side is antisymmetric, we have  $\nabla_{(m} \xi_{n)} =0$ and therefore $\xi$ is a Killing vector:
\beq\label{lie on metric}
\mathcal{L}_{\xi} g =0
\eeq
Actually $\xi$ is more than an isometry. By taking $0=\tr[\jc_0 G^D]= \tr[\jc_0 G^D \Gamma^7]$ from \eqref{susyd1}, we obtain
\beq\label{lie on axion-dilaton}
\mathcal{L}_{\xi}  \phi = \mathcal{L}_{\xi}  C_0 =0 
\eeq
and by using the Bianchi identity for $F_1$ we get
\beq\label{lie on F1}
\mathcal{L}_{\xi} F_1 =0 \ . 
\eeq
Moreover, by taking the trace of \eqref{susye2}, we get
\beq\label{lie on warp factor}
\mathcal{L}_{\xi} A=0 \ . 
\eeq
Using that $\tr[\jc_a G^D \Gamma^6]= \tr[\jc_a G^D \Gamma^{67}]=0$, we get
\beq
\jc_a^{mn} (\ast H)_{mn}=0, \quad a=1,2,3
\eeq
\beq\label{aux1}
\jc_a^{mn} (\ast F_3)_{mn}=0, \quad a=1,2,3 
\eeq
By tracing \eqref{susye3} with $\Gamma^6$ we also get that
\beq\label{aux2}
R=0
\eeq
Then, by tracing \eqref{susye3} with $\Gamma^{67}$  and using \eqref{aux1} with $a=3$, we have 
\beq\label{aux3}
R_7=0
\eeq
The power of the warp factor in the norm of the spinors also comes from supersymmetry. By tracing \eqref{derj} for $a=0$, we get
\beq
\partial_m \rho = 
\dfrac{ \ef}{4 \rt} V_{mn} F^n 
-\dfrac{\ef}{4 \rt} \zeta^n (*F_3)_{mn}
\eeq
The right-hand side can be related to the warp factor by tracing \eqref{susye2} with $\Gamma_{m67}$ which yields\footnote{The integration constant is chosen so that it reproduces the standard value of the charge of the spinors, see \eqref{spinor charges}.}
\beq\label{rho dependence}
\partial_m \rho - \rho \partial_m A =0 \quad \Rightarrow \quad \rho=c \, e^{A} 
\eeq
and we chose $c=1/\sqrt{2}$. Let us now show that the Lie derivative along $\xi$  acting on the rest of the fluxes $H,F_3$ and $F_5$ vanishes. By tracing \eqref{derj} for $a=0$ with $\Gamma_{n7}$ and antisymmetrizing over $[mn]$, we get 
\beq\label{aux10}
\nabla_{[m} \zeta^7_{n]} = - \cl{2} \xi^p H_{mnp} 
\quad \Rightarrow \quad 
d (\iota_{\xi} H) = 0 
\eeq
which by the Bianchi identity for $H$ yields

\beq\label{lie on H}
\mathcal{L}_{\xi} H = 0 
\eeq
The situation for $F_3$ is slightly more complicated due to the non-standard Bianchi identity it satisfies. By tracing \eqref{derj} for $a=0$ with $\Gamma_{n}$ and antisymmetrizing over $[mn]$, we get 
\beq
\nabla_{[m} \zeta_{n]}=
-\cl{4}  (*V)_{pq[m} H_{n]}^{\,\,\,\,\,pq}
-\dfrac{\ef}{4} \xi^p F_{mnp}
-\cl{2\rt} \rho \ef (*F_3)_{mn}
\eeq
We eliminate the H-term using $0=\tr[\jc_0 G^D \Gamma_{mn67}]$ from \eqref{susyd} and we get
\beq\label{aux7}
d \zeta = d\phi \wedge \zeta 
- \ef F_1 \wedge \zeta^7 
-2\ef \iota_{\xi} F_3 
\eeq
Taking the exterior derivative of this expression, replacing again $\iota_{\xi}F_3$ from \eqref{aux7} and using \eqref{aux10}, we get
\beq\label{aux8}
d\iota_{\xi} F_3 + F_1 \wedge \iota_{\xi} H = 0
\eeq
The second term is equal to $\iota_{\xi} d F_3$ as can be seen from the RR Bianchi identities $dF_1 =0 $ and $dF_3 = H \wedge F_1$. Thus, \eqref{aux8} becomes simply
\beq\label{lie on F3}
\mathcal{L}_{\xi} F_3 = 0 
\eeq
In order to compute the the Lie derivative along $\xi$ on $F_5$, we first need $\mathcal{L}_{\xi} \jc_3^7$. By tracing \eqref{derj} with $\Gamma_7$, we get for $a=1,2,3$
\beq
\partial_m \jc_a^7 = 
-\cl{4} \jc_a^{np67} H_{mnp} 
+ \dfrac{i \ef}{8} \jc_a^{np6} F_{mnp} 
+\dfrac{i \ef}{4} \jc^a_{mp} F^p
\eeq
and using $0= \tr[\jc_a G^D \Gamma^{m6}]$ from \eqref{susyd}, we get
\beq\label{aux5}
\partial_m \jc_a^7 = 
\jc_a^7 \partial_m \phi
+ \dfrac{3i \ef}{8} \jc_a^{np6} F_{mnp} 
-\dfrac{3 i \ef}{4} \jc^a_{mp} F^p
\eeq
If we trace \eqref{susye3} with $\Gamma^{m6}$ and replace in the above equation for $a=3$, we get
\beq\label{aux6}
\partial_m \jc_3^7 = \jc_3^7 \partial_m (\phi -3A) \quad \Rightarrow \quad \mathcal{L}_{\xi} \jc_3^7 =0 
\eeq
where \eqref{lie on axion-dilaton} and \eqref{lie on warp factor} were used. Now, it is easy to compute $\mathcal{L}_{\xi} F_5$. Taking the trace of \eqref{susye2} with $\Gamma_7$ and using \eqref{rho dependence} gives
\beq\label{aux9}
m \jc_3^7= - \dfrac{e^{\phi +2A}}{2 \rt} ( \ast F_5 )
\eeq
Taking the Lie derivative along $\xi$ on both sides and using \eqref{lie on metric}, \eqref{lie on axion-dilaton}, \eqref{lie on warp factor} and \eqref{aux6}, we get
\beq\label{lie on F5}
\mathcal{L}_{\xi} F_5 =0 
\eeq
Finally, let us also state another relation which will be useful later. This is easily derived by tracing \eqref{derj} for $a=0$ with $\Gamma_{mn7}$ and eliminating the H-term using $0=\text{Tr} [ \jc_0 G^D \Gamma_{n6}]$. We get
\beq\label{aux11}
\nabla^m V_{mn} =
V_{mn} \partial^m \phi
- \ef \zeta^m_7 (\ast F_3)_{mn} 
+ \zeta^m (\ast H)_{mn} 
- \xi_n (\ast F_5)
\eeq 

\noindent \textbf{The spinor charges}
\vspace*{.2cm}

Here, we compute the charge $q$ of the spinors $\chi_i$ under the U(1) generated by the Killing vector  $\xi$. Actually, it turns out that it is more convenient to compute first $2q$, i.e. the charge of some charged spinor bilinear (we choose $\jc_+^7$), and then divide by 2.  In order to do that, we first need to derive some identities. Multiplying \eqref{fierz 4} with $(\jc_a \Gamma_7)_{\beta \alpha} \jc_0^{\delta \gamma}$ and using $\jc_0 \jc_a= 2 \rho \jc_a$, we get for $a=1,2,3$
\beq\label{JFierz}
\jc_a^{ab} \tr[\jc_0\Gamma_{ab7}] = - 16 \tr[\jc_0 \jc_a \Gamma_7]+8 \rho \jc_a^7  = -24 \rho \jc_a^7 
\eeq
Actually, we can prove a stronger identity by rewriting this in terms of the 5-dimensional spinors $\chi_i$, for which we use  \eqref{charged bilinears}. We will need 
\begin{subequations}
\beq
\jc_+^7 = 4i C_5^{\alpha \beta} \chi^1_{\alpha} \chi^2_{\beta}
\eeq
\beq
\jc_+^{m6}= 4 (\gamma^m)^{\alpha \beta} \chi^1_{\alpha} \chi^2_{\beta}
\eeq
\beq
\jc_+^{mn}= - 2 (\gamma^{mn})^{\alpha \beta} (\chi^1_{\alpha} \chi^1_{\beta} +\chi^2_{\alpha} \chi^2_{\beta} )
\eeq
\end{subequations}
and (see \eqref{neutral0 bilinears})  
\begin{subequations}
\beq
\xi_m= \cl{ \rt} \gamma_m^{\alpha \beta} (\chi^{1c}_{\alpha} \chi^1_{\beta} +\chi^{2c}_{\alpha} \chi^2_{\beta})
\eeq
\beq
V_{mn}= \cl{ \rt} \gamma_{mn}^{\alpha \beta} (\chi^{1c}_{\alpha} \chi^2_{\beta} -\chi^{2c}_{\alpha} \chi^1_{\beta}) \ .
\eeq
\end{subequations}
Using \eqref{fierz 5} and the symmetry properties for gamma matrices in five dimensions, we can show
\beq
V_{mn} \jc_+^{mn} =4 \xi_m \jc_+^{m6}
\eeq
Combining this with \eqref{JFierz} for $a=+$ and using \eqref{rho dependence} we get
\beq\label{aux4}
\xi_m \jc_+^{m6} = -i e^A \jc_+^7
\eeq
Now, we are ready to see how supersymmetry determines the spinor charges.  If we trace \eqref{susye1} with $\Gamma^{m6}$ and replace in \eqref{aux5} for $a=\pm$, we get
\beq
\partial_m \jc_{\pm}^7 =
\jc_{\pm}^7 \partial_m (\phi-3A) 
\mp 3m e^{-A} \jc_{\pm}^{m6}
\eeq
If we contract with $\xi^m$, the first term drops out due to \eqref{lie on axion-dilaton} and \eqref{lie on warp factor}. For the second term, we get using \eqref{aux4}
\beq
\mathcal{L}_{\xi} \jc_{+}^7 =
3 i  m  \jc_+^7
\eeq
and therefore the charges of the spinors $\chi_i$ are 
\beq\label{spinor charges}
q=  \dfrac{3im}{2}
\eeq

\subsection{M-theory}
\label{app:xsi m theory}

The Killing vector in M-theory is the bilinear  \eqref{Killing definition m theory}. This is indeed Killing since \eqref{derj m-theory} yields 
\beq\label{Killingproof m theory}
\nabla_a \xi_b = - \cl{6} \mtf_{abcd} V^{cd} - \dfrac{1}{3\rt} \rho (\star \mtf)_{ab} 
\eeq
and the right-hand side is antisymmetric in $a$ and $b$. Therefore
\beq
\mathcal{L}_{\xi}g =0
\eeq
The trace of \eqref{susye2 m-theory} immediately gives
\beq
\xi^a \partial_a A = \mathcal{L}_{\xi} A=0
\eeq
Finally, we can compute $dV$ by using \eqref{derj m-theory} for $\jc_0$ to get
\beq\label{aux1 m theory}
dV=\iota_{\xi} \mtf_4 \qquad \Longrightarrow \qquad \mathcal{L}_{\xi} \mtf_4 = 0
\eeq
where the Bianchi identity for $\mtf_4$ was used. We see that similarly to the type IIB case, $\xi$ generates a symmetry of the full bosonic sector of the theory.

Let us also derive the warp factor dependence of the normalization of the spinors given by $\theta^{*\alpha}_i \theta_{j,\alpha}= 2 \rho \ \delta_{ij}$. Taking the trace of \eqref{derj m-theory} for $a=0$ and eliminating $\mtf$ 
by taking the trace of \eqref{susye2 m-theory}, we find
\beq\label{rho dependence m theory}
\partial_m \rho - \rho \partial_m A=0 
\quad \Rightarrow \quad \rho = \dfrac{e^{A}}{\rt}
\eeq
where we have chosen the integration constant in the same way as for the IIB case.

\noindent Another useful relation is found by tracing \eqref{derj m-theory} with $\Gamma^{a}$, which yields
\beq\label{aux2 m theory}
\nabla_a \zeta^a = \cl{2} (\star \mtf)_{ab} V^{ab} 
\eeq

Finally let us mention that the M-theory spinor has also definite charge under the action of $\xi$, i.e.
\beq
\mathcal{L}_{\xi} \theta = q \, \theta
\eeq
Matching our conventions with those of \cite{Ashmore:2016qvs}, we find that 
\beq\label{spinor charge m theory}
q=-\dfrac{3im}{2}
\eeq

\section{The moment map for $J_a$}
\label{The moment map for J}

\subsection{Type IIB}
\label{The moment map for J type IIB}

In this section, we prove Eq. \eqref{Hint}, which says that the moment map for the action of a generalized diffeomorphism is related to the dual vector associated to $\kc$ (given by the cubic invariant of \sx $c(\kc, \kc ,V)$). As explained in the main text, this condition can be written  in terms of the twisted moment map density $\mm_a$ which is given by \eqref{twisted moment map density} and we rewrite here for convenience:
\beq\label{moment map appendix}
\mm_a=  
[\nm \jcr_a, \D^m] 
+ (\alg{\nm \mu}{\jcr_a} \,  \D^m)
+\tr [\jcr_a G^{IS}_m] \D^m
-\tr [(\nm \mu) \jcr_a)]\D^m
\eeq
where the second term means the action of $\alg{\nm \mu}{\jcr_a}$ on $\D^m$ while in the rest of the terms $\D^m$ is understood as an element of Cliff(6) and is given by $\D^m=\frac{ie^{2 \phi/3}}{2\rt}  \Gamma^{m67}$. 

Let us  compute the various terms in the above expression. The first term is computed by using \eqref{derj} for $a=1,2,3$. We give the result as a Clifford expansion
\begin{align}\label{moment(1)}
\nonumber  [\nm \jcr_a, \Gamma^{m67}] 
&= \Big[
\cl{16} \jcr^a_{mnpq7} H^{npq} 
-\dfrac{i \ef}{8} \jcr^a_{mn6}F^n 
-\cl{2} \jcr^a_{mn67} \partial^n A
+\cl{2} \jcr^a_{mn67} \partial^n\phi
\Big] \Gamma^{m} \\ \nonumber
&+  \Big[
\cl{8} \jcr_a^{np} H_{mnp} 
+\dfrac{i \ef}{8} \jcr^a_{mn67}F^n 
-\dfrac{i \ef}{48} \jcr^a_{mnpq7}F^{npq}
+\cl{2} \jcr^a_{mn6} \partial^n A
-\cl{2} \jcr^a_{mn6} \partial^n\phi
\Big] \Gamma^{m7} \\ \nonumber
&+  \Big[
\cl{16} \jcr^a_{pqm} H_{n}^{\,\,\,pq} 
+\dfrac{i \ef}{16} \jcr^a_{mnp67}F^p 
-\dfrac{\ef}{16} \jcr^a_{mn} (\ast F_5)
+\cl{2} \jcr^a_{m6} \partial_n A
-\cl{2} \jcr^a_{m6} \partial_n\phi
\Big] \Gamma^{mn7} \\
&+  \Big[
-\cl{8} \jcr^a_{np6} H^{mnp} 
-\dfrac{i \ef}{4} \jcr^a_7 F_m 
-\dfrac{\ef}{4} \jcr^a_{m6} (\ast F_5)
-\cl{2} \jcr^a_{mn} \partial^n A
+\cl{2} \jcr^a_{mn} \partial^n\phi
\Big] \Gamma^{m67}
\end{align}
where the derivatives of the dilaton and the warp factor appear as a result of the rescalings \eqref{jk calligraphic rescaling}.
The second and the fourth term in \eqref{moment map appendix} are those that ``twist'' the moment map density. If we consider them separately they are not gauge invariant, however, their sum is, as it projects onto the fluxes. These terms are computed as follows. For the second term, it is more convenient to use the $SL(6) \times SL(2)$ basis. We first insert \eqref{embedgauge} and the $SL(6) \times SL(2)$ components of $\jcr_a$\footnote{These can be easily found using \eqref{usp8togl5}.} in \eqref{e6alg}. We then use the resulting expression in \eqref{e6dv} to compute the action on $\D^m$ and finally we transform it to the \usp basis using  \eqref{e6f2}. For the fourth term in \eqref{moment map appendix}, we first transform $\nm \mu$ to the \usp basis using \eqref{gl5tousp8a} (exploiting the fact that the $\jc_a$ do not have a \rep{42} component) and then use \eqref{killusp8}. The combined result of these two terms is then\footnote{Here, we mean $\alg{\nm \mu}{\tilde{\jc_a}} \Gamma^{m67}=(\frac{ie^{2 \phi/3}}{2\rt})^{-1}\alg{\nm \mu}{\tilde{\jc_a}}\D^m $.}
\begin{align}\label{moment(2+4)}
\nonumber \alg{\nm \mu}{\tilde{\jc_a}} \Gamma^{m67}
-\tr[\nm \mu \jcr_a] \Gamma^{m67}
&=  \Big[
-\dfrac{1}{24} \jcr^a_{mnpq7} H^{npq} 
-\cl{4} \jcr^a_{mn67} \partial^n \phi
\Big] \Gamma^{m} \\ \nonumber
&+  \Big[
\dfrac{i\ef}{24} \jcr^a_{mnpq7} F^{npq}
+\dfrac{i \ef}{4} \jcr^a_{mn67} F^n
\Big] \Gamma^{m7} \\ \nonumber
&+  \Big[
\dfrac{1}{4} \jcr^a_{m6} \partial_n \phi 
\Big] \Gamma^{mn7} \\ 
&+ \Big[
\dfrac{1}{8} \jcr^a_{np6} H_m^{\,\,\,np} 
-\dfrac{i \ef}{8} \jcr^a_{np67} F_m^{\,\,\,np} 
+\dfrac{\ef}{4} \jcr^a_{m6} (\ast F_5)
-\dfrac{i \ef}{4} \jcr^a_7 F_m
\Big] \Gamma^{m67}
\end{align}
Finally, the third term in \eqref{moment map appendix} is computed directly from \eqref{derj symmetric} and the result reads
\beq\label{moment(3)}
\tr[\jcr_a G^{IS}_m] \Gamma^{m67} =
 \Big[
-\cl{8} \jcr_a^{np6} H_{mnp} 
+\dfrac{i \ef}{8} \jcr_a^7 F_m 
+\dfrac{i \ef}{16} \jcr_a^{np67} F_{mnp}
-\dfrac{\ef}{8} \jcr^a_{m6} (\ast F_5)
\Big] \Gamma^{m67}
\eeq

When adding \eqref{moment(1)}, \eqref{moment(2+4)} and \eqref{moment(3)}, the various terms organize themselves as coefficients of a Cliff(6) expansion. In the next step, we eliminate the H-field using the dilatino equation \eqref{susyd} by taking appropriate traces. More specifically,  we use $\tr{[\jc_a G_d \Gamma^m]}=0$ for the $\Gamma^m$ terms, $\tr{[\jc_a G_d \Gamma^{m7}]}=0$ for the $\Gamma^{m7}$ terms, $\tr{[\jc_a G_d\Gamma^{mn7}]}=0$ for the $\Gamma^{mn7}$ terms and $\tr{[\jc_a G_d\Gamma^{m67}]}=0$ for the $\Gamma^{m67}$ terms. The result is
\begin{align}
\nonumber \Big(\frac{ie^{2 \phi/3}}{2\rt}\Big)^{-1} \mm_a
&= \Big[
\dfrac{i \ef}{8} \jcr^a_{mn6}F^n 
-\dfrac{i \ef}{16} \jcr_a^{np} F_{mnp} 
-\cl{2} \jcr^a_{mn67} \partial^n A
\Big] \Gamma^{m} \\ \nonumber
&+ \Big[
-\dfrac{i \ef}{8} \jcr^a_{mn67}F^n 
-\dfrac{i \ef}{48} \jcr^a_{mnpq7}F^{npq}
+\cl{2} \jcr^a_{mn6} \partial^n A
\Big] \Gamma^{m7} \\ \nonumber
&+  \Big[
-\dfrac{i \ef}{16} \jcr^a_{mnp67}F^p
-\dfrac{i\ef}{16}  \jcr^a_{pqm7} F_{n}^{\,\,\,pq}  
-\dfrac{\ef}{16} \jcr^a_{mn} (\ast F_5)
+\cl{2} \jcr^a_{m6} \partial_n A
\Big] \Gamma^{mn7} \\
&+  \Big[
\dfrac{i \ef}{8} \jcr^a_7 F_m 
+\dfrac{i \ef}{16} \jcr_a^{np67} F_{mnp} 
-\dfrac{\ef}{8} \jcr^a_{m6} (\ast F_5)
-\cl{4} \jcr^a_{mn} \partial^n\phi -\cl{2} \jcr^a_{mn} \partial^n A
\Big] \Gamma^{m67}
\end{align}

For $a=3$ we can find the relation between this and ${\cal K}$ by using the external gravitino equation \eqref{susye3}. Reading off the $\Gamma^m, \Gamma^{m7}, \Gamma^{mn7}$ and $\Gamma^{m67}$ components of this equation, we see that the right-hand sides are exactly the brackets appearing in the above equation. Thus
\begin{align}
\nonumber \mm_3
&= -i \dfrac{m e^{A - 4 \phi/3}}{2} 
\Big[\zeta_m \Gamma^m +i \zeta^7_m \Gamma^{m7} 
+\dfrac{i}{2} V_{mn} \Gamma^{mn7} +i \xi_m \Gamma^{m67} \Big] \\ 
&=-2 i m \rho e^{-4 \phi/3} \kc
\end{align}
where in the last step we used \eqref{k expansion}. Following the same procedure for $a=\pm$ and using this time \eqref{susye1}, we get 
\beq
\mm_\pm=0 
\eeq
These are exactly the conditions \eqref{twistmom computed} which in turn imply the $\tilde {J}_a$ integrability condition \eqref{Hint}.

\subsection{M-theory}
\label{The moment map for J m theory}

In this section, we will present the calculation leading to the integrability condition for the $J_a$ for M-theory compactifications. The methodology is similar to the one for IIB described in the previous subsection. However the details are different due to the different \sx embedding of the derivative and the gauge field in M-theory (Eqs. \eqref{embedder m-theory} and \ref{embedgauge m-theory}). The general expression for the moment map density \eqref{twistmom} now reads\footnote{As in the main text, we omit the $SU(2)$ index $a$ with the understanding that $\jcr=\jcr_{\pm},\jcr_3$.}
\beq
\mm= 
[\nabla_a \jcr, \D^a] 
+ (\alg{\nabla_a \mu}{\jcr} \, \D^a) 
+\tr [\jcr G^{IS}_a]\D^a
-\tr [(\nabla_a \mu) \jcr)]\D^a
\eeq
where now 
\beq \label{vM}
\D^a = \frac{i}{2\rt} \Gamma^{a7}
\eeq
and $G^{I}_a$ is given by \eqref{derj m-theory sym-asym}.

The various terms are computed in exactly the same way as in type IIB so we just give the results here. The first term reads
\begin{align}\label{moment(1) m theory}
\nonumber  [\nabla_a \jcr, \Gamma^{a7}] 
&= \Big[
\cl{72} \jcr^{bcd7} \mtf_{abcd} 
-\cl{2}   \jcr^{7} \partial_a A
\Big] \Gamma^{a} + \\ \nonumber 
&+\Big[
-\cl{36} \jcr^{bcd} \mtf_{abcd}
-\cl{2} \jcr_{ab} \partial^b  A
\Big] \Gamma^{a7} \\
&+  \Big[
\dfrac{i}{6} \jcr^{7} (\star \mtf)_{ab} 
+\cl{48} \jcr^{cd} \mtf_{abcd}
-\cl{4} \jcr_{abc} \partial^c  A
\Big] \Gamma^{ab7} 
\end{align}
while the sum of the second and the fourth is simply
\begin{align}\label{moment(2+4) m theory}
\alg{\nabla_a \mu}{\jcr}\Gamma^{a7}
-\tr[\nabla_a \mu \jcr] \Gamma^{a7}
=\Big[
-\dfrac{i}{8} \jcr^7 (\star \mtf)_{ab}
\Big] \Gamma^{ab7} 
\end{align}
and the third gives
\beq\label{moment(3) m theory}
\tr[\jcr G^I_a] \Gamma^{a7} =
\Big[
-\cl{36} \jcr^{bcd} \mtf_{abcd} 
\Big] \Gamma^{a7}
\eeq

For $\mm=\mm_{\pm}$, we see that the sum of \eqref{moment(1) m theory}, \eqref{moment(2+4) m theory} and \eqref{moment(3) m theory} vanishes by virtue of  \eqref{susye1 m-theory}\footnote{By taking the trace with $\Gamma^{a}$, $\Gamma^{a7}$ and $\Gamma^{ab7}$.}. Thus
\beq
\mm_{\pm} =0 
\eeq
For $\mm=\mm_3$, we follow the same procedure but this time using \eqref{susye3 m-theory}. The result is 
\begin{align}
\nonumber \mm_3
&= -\dfrac{im e ^A}{2}
\Big[\zeta_a \Gamma^a +i \xi_{a} \Gamma^{a7} 
+\dfrac{i}{2} V_{ab} \Gamma^{ab7} \Big] \\ 
&=-2im \rho \kc
\end{align}
where we used \eqref{k expansion m theory}. We this verify the M-theory moment map equation \eqref{Hint} where the rescaled structures are those of \eqref{V and H rescaling M}, are  as in type IIB $\lambda_1=\lambda_2=0$, and $\lambda_3 = -2im$.

\section{The Dorfman derivative along $K$}
\label{The Dorfman derivative along K}

\subsection{Type IIB}
\label{The Dorfman derivative along K type IIB}

The Dorfman derivative is a generalization of the usual Lie derivative for ``generalized flows'' parametrized by the \sx vector $K$. Here we show that the background is invariant under this flow. 

The embedding of the derivative in the \sx object $D$, Eq. \eqref{DV}, picks a particular direction $\D$ in the space of generalized vectors. We start by showing Eq. (\ref{dorf=lie}), namely the fact that the (twisted) Dorfman derivative actually reduces to the Lie derivative along this direction.

As explained in the main text (see \eqref{splitA} and the discussion after that), the twisted Dorfman derivative can be split into a differential piece which is just the directional derivative along the Killing vector $\xi$, given in \eqref{Killingdef}, namely
\beq\label{dorf36(1)}
(\kcr \cdot \D^m) \nabla_m = 
\xi^m \nm
\eeq 
and an algebraic piece $\mathcal{A}$ in the adjoint of \sx. We show that $\mathcal{A}$ satisfies the equations in \eqref{dorf=lie algebraic}.  We start with the \rep{36} piece which according to \eqref{twistd} reads 
\beq\label{dorf36}
\dab{\mathcal{A}}=
(\kcr \cdot \D^m) \dab{\nabla_m \mu } 
- \dab{\big[\nabla_m \kcr, \D^m \big]}
- \dab{\big[(\nabla_m \mu )\kcr, \D^m \big]}
\eeq
where the commutators are just matrix commutators, $\dab{\nm \mu}$ in the first term is just the derivative of the \rep{36} piece of $\mu$ interpreted as a Cliff(6) element, $\dab{((\nabla_m \mu )\kcr)}$ is the standard action\footnote{This term has contributions from both the \rep{36} and the \rep{42} components of $\mu$.} of \sx on the fundamental and \eqref{adjprojusp8a} was used for the projection in the adjoint.

The first and the third term in \eqref{dorf36} twist the Dorfman derivative, so we are computing them together\footnote{Similarly to the moment map equation described in the previous section, each of these terms is not gauge invariant but their sum is.}. $\nm\mu$ is computed just by inserting  (\ref{embedgauge}), in \eqref{gl5tousp8a} while we compete  $(\nabla_m \mu) \kcr$ using \eqref{e6v} and then use \eqref{e6f1} to transform that to the \usp basis. The result is  
\begin{align}\label{dorf36(2+4)}
\nonumber (\kcr \cdot \D^m)  \nabla_m \mu- \big[(\nabla_m \mu )\kcr, \D^m \big]
&=  \Big[
- \cl{6} \xi_m \partial_n \phi
\Big] \Gamma^{mn} \\ \nonumber
&+  \Big[
-\dfrac{1}{4} \zeta^{p} (\ast H)_{np} + \dfrac{\ef}{4} \zeta_7^p (\ast F_3)_{np} - \dfrac{\ef}{4} \xi_n (\ast F_5) + \dfrac{1}{12} V_{mn} \partial^m \phi
\Big] \Gamma^{n6} \\ \nonumber
&+ \Big[
-\dfrac{1}{8} \xi^p H_{mnp} 
 + \dfrac{1}{6} \zeta^7_m \partial_n \phi
\Big] \Gamma^{mn6} \\ 
&+ \Big[
 \dfrac{i\ef}{8} \xi^p F_{mnp} 
 + \dfrac{i}{12} \zeta_{m} \partial_n \phi - \dfrac{i \ef}{4} \zeta^7_{m} F_n
\Big] \Gamma^{mn67} 
\end{align}
where we have expressed the result in terms of the spinor bilinears $\xi, \zeta,\zeta_7$ and $V$ defined in \eqref{k expansion}. 
Finally, the  second term in \eqref{dorf36}  is
easily computed by using \eqref{k expansion}:

\begin{align}\label{dorf36(3)}
\nonumber - \big[\nabla_m \kcr, \D^m \big]
&=  \Big[ 
\cl{4} \nm \xi_n + \cl{6} \xi_m \partial_n \phi
\Big] \Gamma^{mn} \\ \nonumber
&+  \Big[
-\cl{4} \nabla^m V_{mn} +\cl{6} V_{mn} \partial^m \phi 
\Big] \Gamma^{n6} \\ \nonumber
&+  \Big[
-\cl{4} \nabla_m \zeta^7_{n} -\cl{6} \zeta^7_{m} \partial_n \phi 
\Big] \Gamma^{mn6} \\
&+   \Big[
\dfrac{i}{4} \nabla_m \zeta_{n} +\dfrac{i}{6} \zeta_{m} \partial_n \phi 
\Big] \Gamma^{mn67} 
\end{align}
where the derivatives of the dilaton appear due to the rescaling of $\kcr$ given in \eqref{V and H rescaling}.

Collecting the pieces together, i.e. adding  \eqref{dorf36(2+4)} and \eqref{dorf36(3)}, we easily see that the terms proportional to $\Gamma^{n6}$ cancel out due to \eqref{aux11}, those proportional to $\Gamma^{mn6}$ due to \eqref{aux10} and those proportional to $\Gamma^{mn67}$ due to \eqref{aux7}. The remaining terms in \eqref{dorf36(1)} are the sum of the first lines of \eqref{dorf36(2+4)} and \eqref{dorf36(3)} which is simply
\beq
\mathcal{A}|_{\Rep{36}} = \cl{4} (\nm \xi_n) \Gamma^{mn} \ .
\eeq
This is exactly the first equation in \eqref{dorf=lie algebraic}. Let us now look at $\mathcal{A}|_{\mathbf{42}}$, given by 
\beq\label{dorf42}
\mathcal{A}|_{\alpha \beta \gamma \delta}=
(\kcr \cdot \D^m)  \nabla_m \mu_{\alpha \beta \gamma \delta}
-( \D^m \times \nabla_m \kcr )_{\alpha \beta \gamma \delta} 
-(\D^m \times (\nabla_m \mu)\kcr )_{\alpha \beta \gamma \delta}  
\eeq
where the \rep{42} piece of the adjoint projection is given in \eqref{adjprojusp8b}.
The first term  is computed by inserting \eqref{embedgauge} into \eqref{gl5tousp8b} while the third by using \eqref{adjprojusp8b}. Using Fierz identities from appendix \ref{Spinor conventions}, we get for the sum of these two terms
\begin{align}\label{dorf42(2+4)}
\nonumber \Big[
(\kcr \cdot \D^m)  \nabla_m \mu -(\D^m \times (\nabla_m \mu)\kcr )
\Big]_{\alpha \beta \gamma \delta} 
&= \Big[
 \cl{2} \xi_m \partial_n \phi
\Big] \asym{\Gamma^{m67}}{\Gamma^{n67}}\\ \nonumber
&+ \Big[
\dfrac{3}{4} \zeta^p (\ast H)_{np}
-\dfrac{3 \ef}{4} \zeta^p_7 (\ast F_3)_{np} \\ \nonumber
&\qquad+\dfrac{3\ef}{4} \xi_n (\ast F_5) 
-\cl{4} V_{mn} \partial^m \phi
\Big] \asym{\Gamma^{n}}{\Gamma^{6}} \\ \nonumber
&+ \Big[
\dfrac{3i \ef}{8}\xi^p F_{mnp} 
-\dfrac{3i \ef}{4}\zeta^7_m F_n 
+\dfrac{i}{4} \zeta_m \partial_n \phi 
\Big] \asym{\Gamma^{mn7}}{\Gamma^{6}} \\
&+ \Big[
-\dfrac{3}{8}\xi^p H_{mnp} + \cl{2} \zeta^7_m \partial_n \phi
\Big] \asym{\Gamma^{mn7}}{\Gamma^{67}}
\end{align}
containing only the fluxes. The second term in \eqref{dorf42} is given by inserting \eqref{k expansion} in \eqref{adjprojusp8b} and using again some Fierz identities from appendix \ref{Spinor conventions}:
\begin{align}\label{dorf42(3)}
\nonumber -( \D^m \times \nabla_m \kcr )_{\alpha \beta \gamma \delta} 
&=  \Big[ 
- \cl{2} \xi_m \partial_n \phi
\Big] \asym{\Gamma^{m67}}{\Gamma^{n67}} \\ \nonumber
&+  \Big[
\dfrac{3}{4} \nabla^m V_{mn} - \cl{2} V_{mn} \partial^m \phi 
\Big]\asym{\Gamma^{n}}{\Gamma^{6}} \\ \nonumber
&+  \Big[
-\dfrac{3}{4} \nabla_m \zeta^7_{n} -\cl{2} \zeta^7_{m} \partial_n \phi 
\Big] \asym{\Gamma^{mn7}}{\Gamma^{67}} \\
&+   \Big[
\dfrac{3i}{4} \nabla_m \zeta_{n} +\dfrac{i}{2} \zeta_{m} \partial_n \phi 
\Big]  \asym{\Gamma^{mn7}}{\Gamma^{6}}
\end{align}
If we insert now \eqref{dorf42(2+4)} and \eqref{dorf42(3)} in \eqref{dorf42} and use \eqref{aux10}, \eqref{aux7} and \eqref{aux11}  (as for the \rep{36} component), we get
\beq
\mathcal{A}|_{\mathbf{42}}=0 \ ,
\eeq
which completes thus the proof of \eqref{dorf=lie algebraic}. Combining this with \eqref{dorf36(1)} and the fact that the $\jc_a$ have only a \rep{36} component we arrive at \eqref{dorf=lie} as we explain in the main text.

\subsection{M-theory}
\label{The Dorfman derivative along K m theory}

Let us now perform the same kind of calculation for the M-theory set-up. Although the details are different than in type IIB, the basic procedure to prove that the twisted Dorfman derivative along $\kc$ is equal to the usual Lie derivative along the corresponding Killing vector is actually the same. The differential piece is again the directional derivative along $\xi$\footnote{We recall that $\kcr = \kc$ for the M-theory case.}
\beq\label{dorf36(1) m theory}
(\kc \cdot \D^a) \nabla_a = 
\xi^a \nabla_a
\eeq 
The \rep{36} piece of the operator $\mathcal{A}$  is given by
\beq\label{dorf36 m theory}
\dab{\mathcal{A}} =  
(\kc \cdot \D^a) \dab{\nabla_a \mu} 
- \dab{\big[\nabla_a \kc, \D^a \big]}
- \dab{\big[(\nabla_a \mu )\kc, \D^a \big]}
\eeq
The first term together with the third is
\begin{align}\label{dorf36(2+4) m theory}
(\kc \cdot \D^a)  \nabla_a \mu- \big[(\nabla_a \mu )\kc, \D^a \big]
=  \Big[
\cl{24} \xi^d  G_{abcd}
\Big] \Gamma^{abc}
+  \Big[\dfrac{i}{8} V^{ab} (\star F)_{ab}
\Big] \Gamma_{7} \ .
\end{align}
while the second is
\begin{align}\label{dorf36(3) m theory}
- \big[\nabla_a \kc, \D^{a7}] 
=  \Big[
 \cl{4} \nabla_a \xi_b
\Big] \Gamma^{ab} 
+  \Big[
-\cl{8} \nabla_a V_{bc} 
\Big] \Gamma^{abc} 
+   \Big[
 - \dfrac{i}{4} \nabla_a \zeta^a
\Big] \Gamma_7 \ .
\end{align}
It is straightforward to see using \eqref{aux1 m theory} and \eqref{aux2 m theory} that their sum is just 
\beq
\mathcal{A}|_{\Rep{36}} = \cl{4} (\nabla_a \xi_b) \Gamma^{ab}
\eeq
We finally show that  $\mathcal{A}|_{\mathbf{42}}=0$ also in M-theory. We have
\beq\label{dorf42 m theory}
\mathcal{A}_{\alpha \beta \gamma \delta}=
(\kc \cdot \D^a)  \nabla_a \mu_{\alpha \beta \gamma \delta}
-( \D^a \times \nabla_a \kc )_{\alpha \beta \gamma \delta} 
-(\D^a \times (\nabla_a \mu)\kc )_{\alpha \beta \gamma \delta}  
\eeq
Similarly to type IIB
\begin{align}\label{dorf42(2+4) M theory}
\Big[(\kc \cdot \D^a)  \nabla_a \mu
-(\D^a \times (\nabla_a \mu)\kc )
\Big]_{\alpha \beta \gamma \delta}
&= \Big[\dfrac{i}{16} V^{ab} (\star G)_{ab}\Big] \asym{\Gamma^{c}}{\Gamma^{c7}}
+\Big[-\dfrac{1}{8} \xi^d G_{abcd} 
 \Big] \asym{\Gamma^{a7}}{\Gamma^{bc7}}
\end{align}
where we have used \eqref{fierz 1} and \eqref{fierz 2} to simplify the terms proportional to $V$ and \eqref{fierz 3} for the terms proportional to $\xi$. Using \eqref{k expansion m theory}, we also get
\begin{align}
-\Big[ \D^a \times \nabla_a \kc ]_{\alpha \beta \gamma \delta}
&= \Big[-\dfrac{i}{8} \nabla_a \zeta^a\Big] \asym{\Gamma^{c}}{\Gamma^{c7}}
+\Big[-\dfrac{3}{8} \nabla_{[a} V_{bc]} \Big] \asym{\Gamma^{a7}}{\Gamma^{bc7}}
\end{align}
where again the terms proportional to derivatives of $\zeta$ are absent because of \eqref{fierz 1} \eqref{fierz 2} while due to \eqref{fierz 3} only the exterior derivative of $V$ appears. The sum of \eqref{dorf42(2+4) M theory} and \eqref{dorf42(3)} vanishes using \eqref{aux1 m theory} and \eqref{aux2 m theory}. We thus get
\beq
\mathcal{A}|_{\mathbf{42}}=0
\eeq
and therefore we verify \eqref{dorf=lie} for M-theory as well.

\bibliographystyle{utphys}
\bibliography{Draftnew.bib}

\providecommand{\href}[2]{#2}\begingroup\raggedright\begin{thebibliography}{10}

\bibitem{Gauntlett:2003cy}
J.~P. Gauntlett, D.~Martelli, and D.~Waldram, ``Superstrings with intrinsic
  torsion,'' {\em Phys. Rev.} {\bf D69} (2004) 086002,
  \href{http://xxx.lanl.gov/abs/hep-th/0302158}{{\tt hep-th/0302158}}.

\bibitem{Hitchin:2004ut}
N.~Hitchin, ``Generalized calabi-yau manifolds,'' {\em Quart. J. Math. Oxford
  Ser.} {\bf 54} (2003) 281--308,
  \href{http://xxx.lanl.gov/abs/math.dg/0209099}{{\tt math.dg/0209099}}.

\bibitem{Gualtieri:2003dx}
M.~Gualtieri, ``Generalized complex geometry,''
  \href{http://xxx.lanl.gov/abs/math.dg/0401221}{{\tt math.dg/0401221}}.

\bibitem{Pacheco:2008ps}
P.~P. Pacheco and D.~Waldram, ``{M-theory, exceptional generalised geometry and
  superpotentials},'' {\em JHEP} {\bf 09} (2008) 123,
  \href{http://xxx.lanl.gov/abs/0804.1362}{{\tt 0804.1362}}.

\bibitem{Grana:2009im}
M.~Grana, J.~Louis, A.~Sim, and D.~Waldram, ``{E7(7) formulation of N=2
  backgrounds},'' {\em JHEP} {\bf 07} (2009) 104,
  \href{http://xxx.lanl.gov/abs/0904.2333}{{\tt 0904.2333}}.

\bibitem{Coimbra:2014uxa}
A.~Coimbra, C.~Strickland-Constable, and D.~Waldram, ``{Supersymmetric
  Backgrounds and Generalised Special Holonomy},''
  \href{http://xxx.lanl.gov/abs/1411.5721}{{\tt 1411.5721}}.

\bibitem{Grana:2005tf}
M.~Grana, R.~Minasian, A.~Tomasiello, and M.~Petrini, ``Supersymmetric
  backgrounds from generalized calabi-yau manifolds,'' {\em Fortsch. Phys.}
  {\bf 53} (2005) 885--893.

\bibitem{Grana:2011nb}
M.~Grana and F.~Orsi, ``{N=1 vacua in Exceptional Generalized Geometry},''
  \href{http://xxx.lanl.gov/abs/1105.4855}{{\tt 1105.4855}}.

\bibitem{Lust:2010by}
D.~Lust, P.~Patalong, and D.~Tsimpis, ``{Generalized geometry, calibrations and
  supersymmetry in diverse dimensions},'' {\em JHEP} {\bf 01} (2011) 063,
  \href{http://xxx.lanl.gov/abs/1010.5789}{{\tt 1010.5789}}.

\bibitem{Coimbra:2015nha}
A.~Coimbra and C.~Strickland-Constable, ``{Generalised Structures for
  $\mathcal{N}=1$ AdS Backgrounds},''
  \href{http://xxx.lanl.gov/abs/1504.02465}{{\tt 1504.02465}}.

\bibitem{Ashmore:2015joa}
A.~Ashmore and D.~Waldram, ``{Exceptional Calabi--Yau spaces: the geometry of
  $\mathcal{N}=2$ backgrounds with flux},''
  \href{http://xxx.lanl.gov/abs/1510.00022}{{\tt 1510.00022}}.

\bibitem{Grana:2005sn}
M.~Grana, R.~Minasian, M.~Petrini, and A.~Tomasiello, ``Generalized structures
  of n = 1 vacua,'' {\em JHEP} {\bf 11} (2005) 020,
  \href{http://xxx.lanl.gov/abs/hep-th/0505212}{{\tt hep-th/0505212}}.

\bibitem{Ashmore:2016qvs}
A.~Ashmore, M.~Petrini, and D.~Waldram, ``{The exceptional generalised geometry
  of supersymmetric AdS flux backgrounds},''
  \href{http://xxx.lanl.gov/abs/1602.02158}{{\tt 1602.02158}}.

\bibitem{Louis:2012ux}
J.~Louis, P.~Smyth, and H.~Triendl, ``{Supersymmetric Vacua in N=2
  Supergravity},'' {\em JHEP} {\bf 08} (2012) 039,
  \href{http://xxx.lanl.gov/abs/1204.3893}{{\tt 1204.3893}}.

\bibitem{Coimbra:2011nw}
A.~Coimbra, C.~Strickland-Constable, and D.~Waldram, ``{Supergravity as
  Generalised Geometry I: Type II Theories},'' {\em JHEP} {\bf 11} (2011) 091,
  \href{http://xxx.lanl.gov/abs/1107.1733}{{\tt 1107.1733}}.

\bibitem{Hull:2007zu}
C.~M. Hull, ``{Generalised Geometry for M-Theory},'' {\em JHEP} {\bf 07} (2007)
  079, \href{http://xxx.lanl.gov/abs/hep-th/0701203}{{\tt hep-th/0701203}}.

\bibitem{Coimbra:2011ky}
A.~Coimbra, C.~Strickland-Constable, and D.~Waldram, ``{$E_{d(d)} \times
  \mathbb{R}^+$ Generalised Geometry, Connections and M theory},''
  \href{http://xxx.lanl.gov/abs/1112.3989}{{\tt 1112.3989}}.

\bibitem{Coimbra:2012af}
A.~Coimbra, C.~Strickland-Constable, and D.~Waldram, ``{Supergravity as
  Generalised Geometry II: $E_{d(d)} \times \mathbb{R}^+$ and M theory},'' {\em
  JHEP} {\bf 03} (2014) 019, \href{http://xxx.lanl.gov/abs/1212.1586}{{\tt
  1212.1586}}.

\bibitem{Gauntlett:2005ww}
J.~P. Gauntlett, D.~Martelli, J.~Sparks, and D.~Waldram, ``{Supersymmetric
  AdS(5) solutions of type IIB supergravity},'' {\em Class. Quant. Grav.} {\bf
  23} (2006) 4693--4718, \href{http://xxx.lanl.gov/abs/hep-th/0510125}{{\tt
  hep-th/0510125}}.

\bibitem{Maldacena:2000mw}
J.~M. Maldacena and C.~Nunez, ``{Supergravity description of field theories on
  curved manifolds and a no go theorem},'' {\em Int. J. Mod. Phys.} {\bf A16}
  (2001) 822--855, \href{http://xxx.lanl.gov/abs/hep-th/0007018}{{\tt
  hep-th/0007018}}. [,182(2000)].

\bibitem{Gauntlett:2004zh}
J.~P. Gauntlett, D.~Martelli, J.~Sparks, and D.~Waldram, ``{Supersymmetric
  AdS(5) solutions of M theory},'' {\em Class. Quant. Grav.} {\bf 21} (2004)
  4335--4366, \href{http://xxx.lanl.gov/abs/hep-th/0402153}{{\tt
  hep-th/0402153}}.

\bibitem{Louis:2016qca}
J.~Louis and C.~Muranaka, ``{Moduli spaces of $AdS_5$ vacua in N=2
  supergravity},'' \href{http://xxx.lanl.gov/abs/1601.00482}{{\tt 1601.00482}}.

\bibitem{Bergshoeff:2001pv}
E.~Bergshoeff, R.~Kallosh, T.~Ortin, D.~Roest, and A.~Van~Proeyen, ``{New
  formulations of D = 10 supersymmetry and D8 - O8 domain walls},'' {\em Class.
  Quant. Grav.} {\bf 18} (2001) 3359--3382,
  \href{http://xxx.lanl.gov/abs/hep-th/0103233}{{\tt hep-th/0103233}}.

\bibitem{Charlie}
A.~Coimbra and C.~Strickland-Constable, ``to appear,''.

\bibitem{Ash}
A.~Ashmore, M.~Gabella, M.~Gra\~na, M.~Petrini, and D.~Waldram, ``{Exactly
  marginal deformations from exceptional generalised geometry},''
  \href{http://xxx.lanl.gov/abs/1605.05730}{{\tt 1605.05730}}.

\end{thebibliography}\endgroup

\end{document}